\documentclass[preprint,aps]{revtex4}
\usepackage[latin9]{luainputenc}
\usepackage{float}
\usepackage{amsmath}
\DeclareMathOperator\arctanh{arctanh}
\DeclareMathOperator\arcsinh{arcsinh}
\usepackage{amssymb}
\usepackage{graphicx}
\usepackage{setspace}
\usepackage[unicode=true,pdfusetitle,
 bookmarks=true,bookmarksnumbered=false,bookmarksopen=false,
 breaklinks=false,pdfborder={0 0 1},backref=false,colorlinks=false]
 {hyperref}

\makeatletter

\@ifundefined{textcolor}{}
{%
 \definecolor{BLACK}{gray}{0}
 \definecolor{WHITE}{gray}{1}
 \definecolor{RED}{rgb}{1,0,0}
 \definecolor{GREEN}{rgb}{0,1,0}
 \definecolor{BLUE}{rgb}{0,0,1}
 \definecolor{CYAN}{cmyk}{1,0,0,0}
 \definecolor{MAGENTA}{cmyk}{0,1,0,0}
 \definecolor{YELLOW}{cmyk}{0,0,1,0}
}

\usepackage{amsfonts}
\usepackage{bm}
\usepackage[dvips]{epsfig}
\usepackage[dvips]{graphicx}
\usepackage{float}
\usepackage{dcolumn}
\usepackage{ifpdf}
\usepackage{dcolumn}
\usepackage{multirow}

\newcommand{\be}{\begin{equation}}
\newcommand{\ee}{\end{equation}}
\newcommand{\bes}{\begin{subequations}}
\newcommand{\ees}{\end{subequations}}
\newcommand{\ben}{\begin{eqnarray}}
\newcommand{\een}{\end{eqnarray}}

\makeatother

\begin{document}
\title{Oscillons in hyperbolic models}
 \author{D. Bazeia${^1}$, Adalto R. Gomes$^{2}$, K. Z. Nobrega$^{3}$, Fabiano C. Simas$^{4}$, }
%\href{mailto:name.surname@gmail.com}{name.surname@gmail.com}
\email{[1] {dbazeia@gmail.com}, [2] {argomes.ufma@gmail.com}, [3] {bzuza1@yahoo.com.br }, [4] {simasfc@gmail.com }}
\noaffiliation
\affiliation{
$^1$ Departamento de F\'isica, Universidade Federal da Para\'iba, 58051-970, Jo\~ao Pessoa, PB, Brazil\\
$^2$ Departamento de F\'isica, Universidade Federal do Maranh\~ao (UFMA)
Campus Universit\'ario do Bacanga, 65085-580, S\~ao Lu\'\i s, Maranh\~ao, Brazil\\
$^3$ Departamento de Eletro-Eletr\^onica, Instituto Federal de Educa\c c\~ao, Ci\^encia e
Tecnologia do Maranh\~ao (IFMA), Campus Monte Castelo, 65030-005, S\~ao Lu\'is, Maranh\~ao, Brazil\\
$^4$ Centro de Ci\^encias Agr\'arias e Ambientais-CCAA, Universidade Federal do Maranh\~ao
(UFMA), 65500-000, Chapadinha, Maranh\~ao, Brazil
}
\noaffiliation

\begin{abstract}
In this work we examine kink-antikink collisions in two distinct hyperbolic models. The models depend on a deformation parameter, which controls two main characteristics of the potential with two degenerate minima: the height of the barrier and the values of the minima. In particular, the rest mass of the kinks decreases monotonically as the deformation parameter increases, and we identify the appearance of a gradual suppression of two bounce windows in the kink scattering and the production of long lived oscillons. The two effects are reported in connection to the presence of more than one vibrational state in the stability potential.
\end{abstract}

%\pacs{ 04.60.Kz, 11.27.+d}


\maketitle

%%%%%%%%%%%%%%%%%%%%%%%%%%%%%%%%%%%%%%%%%%%%%%%%%%%%%%%%%%%%%%%%%%%%%%%%%%%%%%%

\section{ Introduction }

Localized structures are important in nonlinear physics. In low and high energy physics, localized structures have been studied in several different contexts \cite{vacha,mastu,daupey}. In high energy physics, in particular, nontrivial localized structures appear as kinks, vortices and monopoles in $(1,1)$, $(2,1)$ and $(3,1)$ spacetime dimensions, respectively \cite{mastu}. In the simplest situation, kinks and antikinks appear in scalar field theories described by a single real scalar field. 

In nonintegrable scalar field theories like the $\phi^4$ model \cite{vacha}, the existence of kinks and antikinks motivates the study of their scattering, that may sometimes lead to surprisingly rich consequences. For instance, when the collision is analyzed as a function of the initial velocity of approximation of the two structures, a complicated structure appears \cite{kudry}, usually connected with the deformation of the field profile and the emission of radiation. However, for larger initial velocities a simple inelastic scattering occurs and the kink-antikink pair retreats from each other. In the richer case with sufficiently small initial velocities, the kink and antikink capture one another, forming a trapped bion state that radiates continuously until being completely annihilated. 

 An intriguing aspect of the collision, observed in particular in the well-known $\phi^4$  model \cite{camschowin,aniolmat,goodhab}, occurs for some windows of intermediate velocities, named two-bounce windows, where the scalar field at the center of mass bounces twice before the pair recedes to infinity. These windows appear in sequence with smaller thickness, accumulating in the border of the one-bounce region.  The same effect was also verified for higher levels of bounce windows, leading to a fractal structure \cite{aniolmat}. The two-bounce windows were interpreted in the Ref. \cite{camschowin} as related to the exchange of energy between the translational and vibrational modes that are present in the model. The $\phi^6$ model is an exception for this mechanism, since the resonant scattering appears if one considers the effect of collective modes produced by the antikink-kink pair \cite{domerosh}. Another counterexample of the mechanism described in the Ref. \cite{camschowin} appeared in \cite{sgno}; there, there are no two-bounce windows even in the  presence of vibrational modes. Moreover, 
we want to add that the richness of the scattering may also be connected with the internal structure of the stability potential that appears in the model, which is related with the potential that defines the model and gives rise to the kinks and antikinks. The sense of this is that the internal modes of the stability potential can provide new windows or resonances that can modify the profile of the collision, leading to novel possibilities of current interest.  

The study of collisions of kink and antikink has gained further attention recently, with the study of polynomial models with one \cite{domerosh,dedekecrsa,roman1,ganilenliz1} and two or more \cite{halromshn,al1,al2,al3} scalar fields, of nonpolynomial models \cite{gankud,sgn,gaaes,bbv2} and of models that support multi-kink
configurations \cite{magasaadmja,maassadm,almasazhdi,gan3}. Moreover, there are investigations on the collision of large relativistic bubbles, that can be treated as that of planar walls, described as kink scattering in $(1,1)$ dimensions \cite{gib}. Kinks were also proposed in buckled graphene nanoribbon \cite{graph1, graph2}, and they also appear as topological excitations in {\it trans}-polyacetilene. For instance, the lattice model of the Su-Schrieffer-Heeger \cite{ssh} also predicted that kinks can propagate as independent entities along the polymer.

Kinks also find interesting applications in ferroelectric materials. As one knows, polynomial and modified sine-Gordon models suffer from a drastic weakness due to the rigidity of some ferroelectric materials. Indeed, in these models the barrier height of the double-well potential cannot be varied as a function of the shape parameter. To address this question, an hyperbolic extension \cite{dk1,dk2} of the $\phi^4$ model was proposed to describe the structural transitions observed in specific materials \cite{dk3}. These models belong to the class of deformed double-well potentials $V(\phi,\mu)$, where $\phi$ corresponds to the order parameter and $\mu$ is a deformation parameter. In general the functions of $\mu$ are introduced at will in the potential, in a phenomenological construction. 

The Calogero model \cite{cal} describes $N$ identical non-relativistic particles in one dimension, having exact soliton solutions in the continuum limit \cite{poly}. A hyperbolic extension of the Calogero model was shown to be integrable even in strong confinement, and presenting  multi-soliton solutions \cite{gon}. $\mathcal{N}=2$ and $\mathcal{N}=4$ supersymmetric  generalizations of the hyperbolic Calogero model were presented in the Ref. \cite{susy}. Hyperbolic models also have been  applied to attain exact solutions for hairy black holes \cite{c}. Moreover, a generalised inverse cosine-hyperbolic potential was considered to describe quintessential inflation \cite{agar}. Tachyon matter cosmology with hyperbolic potentials has also been considered in the Ref. \cite{pour}. 

Motivated by the above investigations, in this work we consider the scattering of kinks in models defined in (1,1) spacetime dimensions, in the presence of hyperbolic potentials. In the next section we consider two different models, which are inspired by the $\phi^4$ model and the Refs. \cite{dk1,dk2}, but we concentrate on collecting results for the kink-antikink collisions and their departure from the standard results obtained with the $\phi^4$ model. An interesting result concerns the production of oscillons, the long-lived and low-amplitude oscillation of the scalar field around the trivial configuration. We finish the work in Sect. III, with some comments and conclusions.

%%%%%%%%%%%%%%%%%%%%%%%%%%%%%%%%%%%%%%%%%%%%%%%%%%%%%%%%%%%%%%%%%%%%%%%%%%%%%%%An interesting result 

\section{Hyperbolic models}

%%%%%%%%%%%%%%%%%%%%%%%%%%%%%%%%%%%%%%%%%%%%%%%%%%%%%%%%%%%%%%%%%%%%%%%%%%%%%%%

We start with the standard action
\begin{equation}
S=\int  dtdx \biggl( \frac12 \partial_\mu\phi\partial^\mu\phi - V(\phi) \biggr),
\end{equation}
where the potential has two minima and a local maximum at the origin. Then we have one topological sector connecting adjacent minima. The equation of motion is given by
\begin{equation}
\frac{\partial^2\phi}{\partial t^2}-\frac{\partial^2\phi}{\partial x^2}+\frac{dV}{d\phi}=0.
\end{equation}
Static kink $\phi_{\bar K}(x)$ and antikink $\phi_{\bar K}=\phi_K(-x)$ are solutions that connect the two sectors of the potential. Perturbing linearly the scalar field around one kink solution as $\phi(x,t)=\phi_K(x)+\eta(x)\cos(\omega t)$ we get a Schr\"odinger-like equation
\be
-\frac{\partial^2\eta}{\partial x^2}+ V_{sch}\,\eta=\omega^2\eta,
\ee
with $V_{sch}(x)=\frac{d^2 V}{d\phi^2}$ being the stability potential. The analysis of the Schr\"odinger-like or stability potential is useful for understanding some aspects of the scattering structure.

 For the numerical solutions of kink-antikink scattering we used a $4^{th}$ order finite-difference method on a grid $N=4096$ nodes and a spatial step $\delta = 0.05$. We fixed $x=\pm x_0$ with $x_0=12$ for the initial symmetric position of the pair and set the grid boundaries at $ x=\pm x_{max}$ with $x_{max}=400$. For the time dependence we used a $6^{th}$ order sympletic integrator method, with a time step $\delta=0.02$.

For solving the equation of motion for kink-antikink scattering we used the following initial conditions
\begin{eqnarray}
\phi(x,0)&=&\phi_K(x+x_0,v,0)-\phi_{K}(x-x_0,-v,0)-\phi_v,\\
\dot\phi(x,0)&=&\dot\phi_K(x+x_0,v,0)-\dot\phi_{K}(x-x_0,-v,0),
\end{eqnarray}
where $\phi_K(x+x_0,v,t)$ means a boost solution for kink and $\phi_v>0$ is one vacuum of the theory (minimum of $V(\phi_v)$). 

%%%%%%%%%%%%%%%%%%%%%%%%%%%%%%%%%%%%%%%%%%%%%%%%%%%%%%%%%%%%%%%%%%%%%%%%%%%%%%%

\subsection{Model 1} \label{secII}

\begin{figure}
		\includegraphics[{angle=0,width=8cm}]{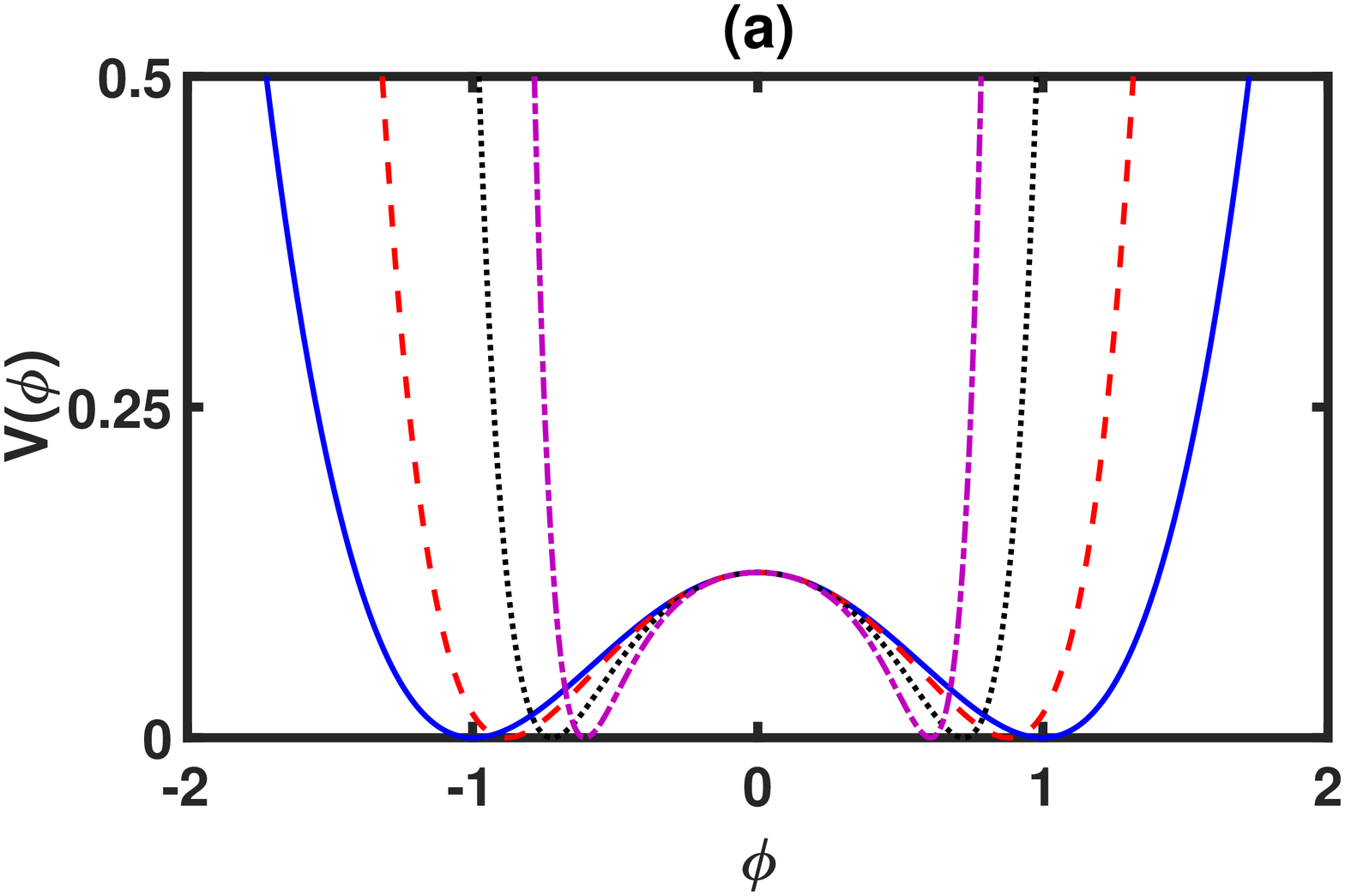}
		\includegraphics[{angle=0,width=8cm}]{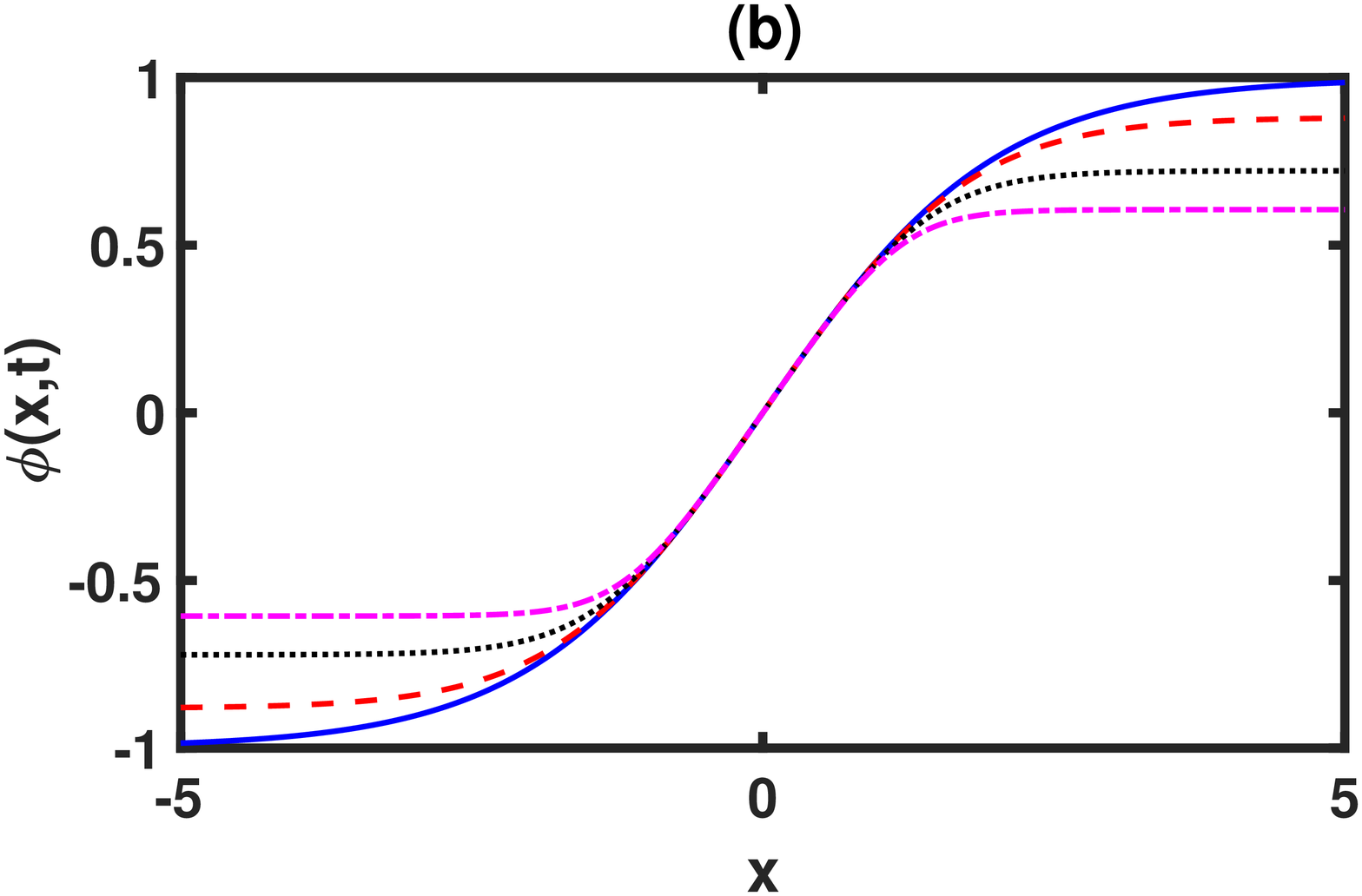}
		\includegraphics[{angle=0,width=8cm}]{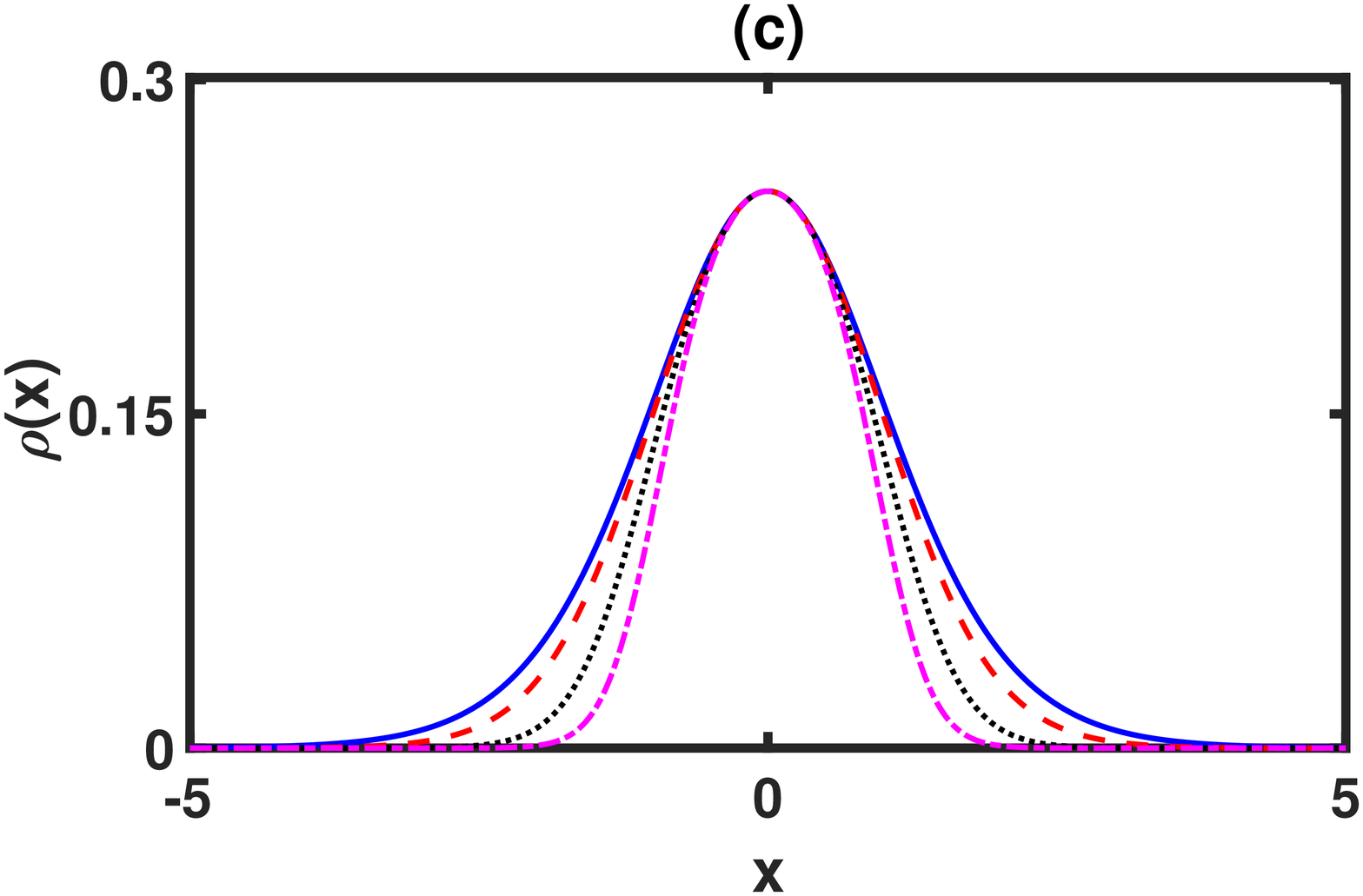}
		\includegraphics[{angle=0,height=5cm,width=8cm}]{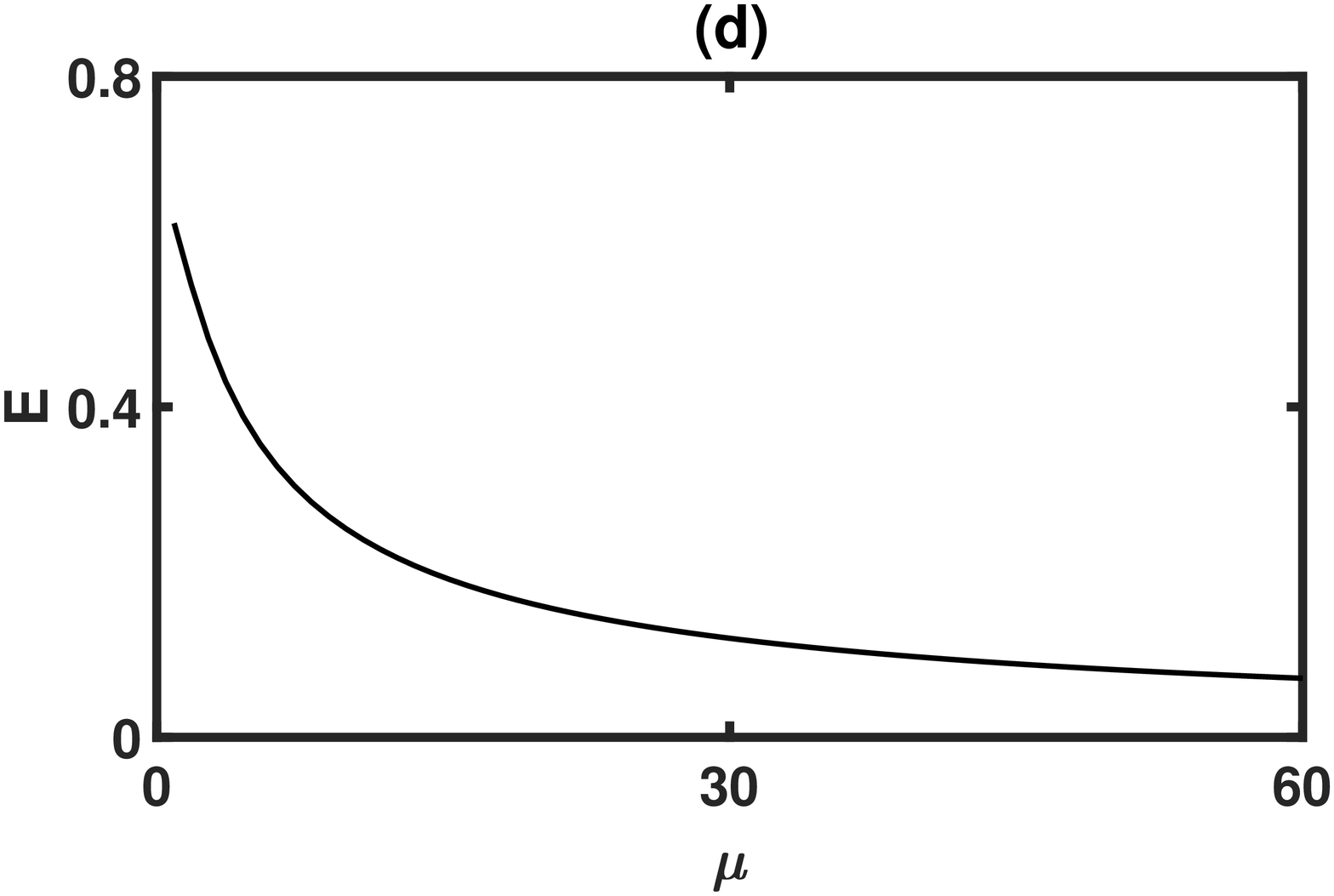}
\caption{Model 1: (a) Potential $V_1(\phi)$, (b) field $\phi(x)$ and (c) energy density $\rho(x)$ for fixed
$\mu=0.1$ (blue solid), $\mu=1$ (red dash), $\mu=2$ (black dotted) and $\mu=3$ (purpple dash-dotted), and (d) the kink energy or rest mass as function of $\mu$.}
\label{V1-phi-dens}
\end{figure}

%%%%%%%%%%%%%%%%%%%%%%%%%%%%%%%%%%%%%%%%%%%%%%%%%%%%%%%%%%%%%%%%%%%%%%%%%%%%%%%
We consider the potential \cite{dk1}
\begin{eqnarray}
V_1(\phi)=\frac18\bigg(\frac{\sinh^2(\mu \phi)}{\mu^2} - 1 \bigg)^2,
\label{pot1}
\end{eqnarray}
where $\mu$ is the deformability parameter. The Fig. \ref{V1-phi-dens}a shows that the potential has two minima in $\phi=\pm(1/\mu)\arcsinh(\mu)$.  In this form, the minima of the potential are variable, but the height of its barrier is the same. In the limit of small values of $\mu$, the model approaches the usual polynomial $\phi^4$ theory with minima at $\phi=\pm1$.

%%%%%%%%%%%%%%%%%%%%%%%%%%%%%%%%%%%%%%%%%%%%%%%%%%%%%%%%%%%%%%%%%%%%%
%%%%%%%%%%%%%%%%%%%%%%%%%%%%%%%%%%%%%%%%%%%%%%%%%%%%%%%%%%%%%%%%%%%%%
 Static kink solution and the minima $\pm\phi_v$ are given by \cite{dk1}     
\begin{eqnarray}
  \phi_K(x) &=& \frac{1}{\mu}\arctanh\Bigg( \frac{\mu \tanh\bigg( \frac{\sqrt{1+\mu^2}x}{2} \bigg)}{\sqrt{1+\mu^2}} \Bigg), \label{sol1-phi}\\
 \phi_v &=& \frac{1}{\mu}\arcsinh(\mu).
    \label{sol1}
\end{eqnarray}
The corresponding energy (rest mass) is given by
\be
E=\int_{-\infty}^{+\infty} \rho(x) dx = \int_{-\infty}^{+\infty} \bigg( \frac12 \Big( \frac{d \phi}{dx} \Big)^2 + V(\phi)  \bigg)  dx.
\ee	
In this model, the potential can be written as
\be
V_1(\phi)=\frac12 \biggl( \frac{dW_1}{d\phi} \biggr)^2
\ee
and so it admits the first order equations
\be
\frac{d\phi}{dx}=\pm \frac{dW_1}{d\phi} = \pm \frac12 \biggl(  \frac{\sinh^2(\mu\phi)}{\mu^2}-1 \biggr).
\ee
The kink in the Eq. (\ref{sol1-phi}) obeys the above equation with the $+$ sign. The antikink  obeys the same equation with the $-$ sign. This is important since both kink and antikink are minimum energy solutions and so they are linearly stable.

The Fig. \ref{V1-phi-dens}b depicts the plots of the scalar field profile $\phi(x)$ for some values of $\mu$. Note that the asymptotic value decreases with $\mu$. The energy density  $\rho(x)$, as shown in the Fig. \ref{V1-phi-dens}c  is a localized function around $x=0$. From the plot one sees that its maximum is fixed, whereas the thickness decreases with the increasing of $\mu$. This means that as $\mu$ increases, the solution becomes more and more localized. In the Fig. \ref{V1-phi-dens}d we see that the kink rest energy decreases with the increasing of $\mu$.

%%%%%%%%%%%%%%%%%%%%%%%%%%%%%%%%%%%%%%%%%%%%%%%%%%%%%%%%%%%%%%%%%%%%%
\begin{figure}
	\includegraphics[{angle=0,width=8cm,height=4cm}]{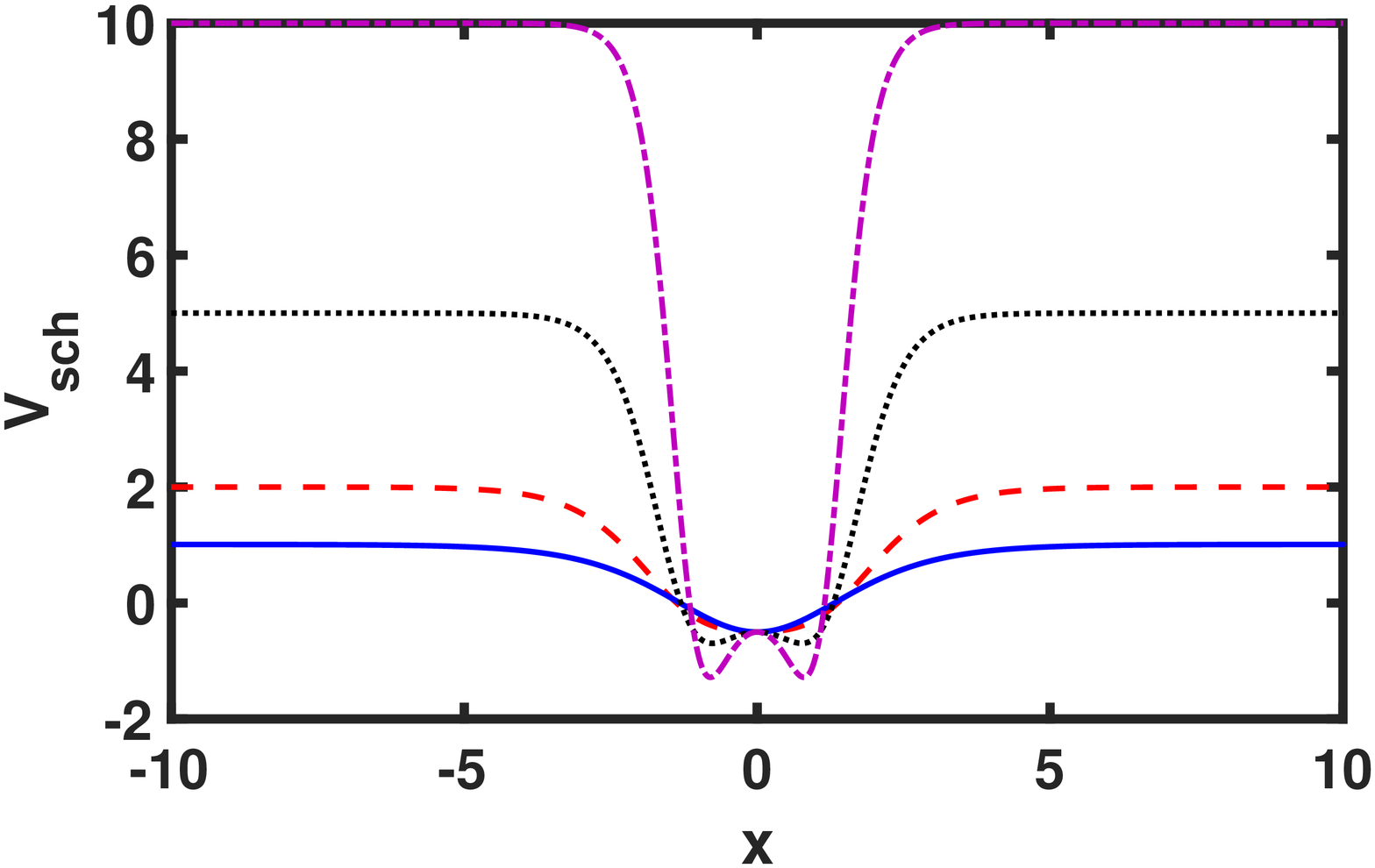}
	\includegraphics[{angle=0,width=8cm,height=4cm}]{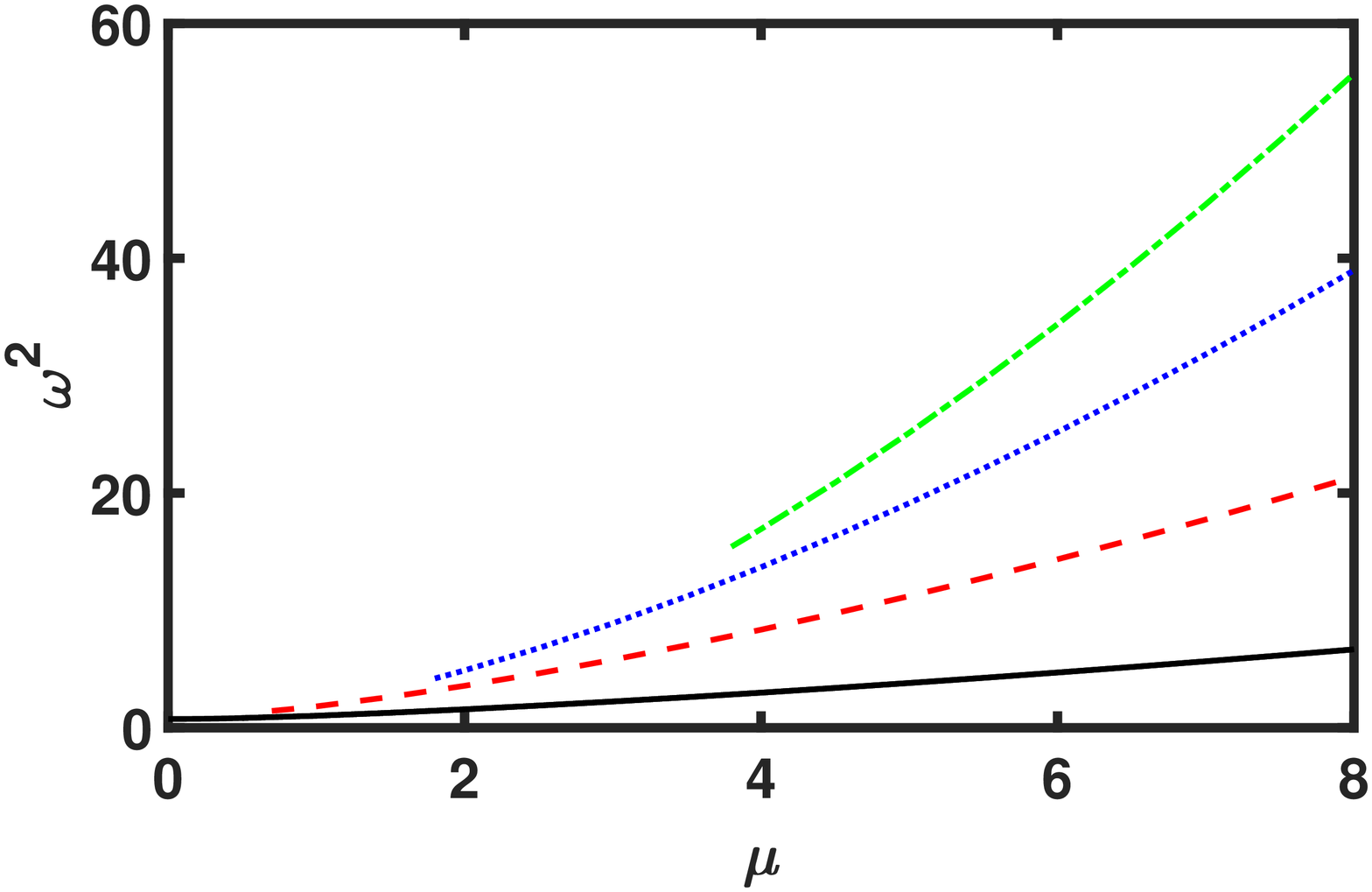}
	\caption{ (Left) The Schr\"odinger-like potential $V_{sch}$ for fixed $\mu=0.1$ (blue solid), $\mu=1$ (red dash), $\mu=2$ (black dotted) and $\mu=3$ (purpple dash-dotted). (Right) The squared frequencies $\omega^2$ of the vibrational states as a function of the parameter $\mu$.}
	\label{V1-sch-wsq}
\end{figure}
%%%%%%%%%%%%%%%%%%%%%%%%%%%%%%%%%%%%%%%%%%%%%%%%%%%%%%%%%%%%%%%%%%%%%

The Schr\"odinger-like or stability potential is given by
\be
V_{sch}(x)= \frac{(\mu^2+1)\bigg[\mu^2 \tanh^4\big(q(x)\big) +3\tanh^2\big(q(x)\big) -\mu^2-1 \bigg] }{2\bigg(\mu^2\tanh^2\big(q(x)\big)-\mu^2-1 \bigg)^2},
\ee
where $q(x)=x\sqrt{\mu^2+1}/2$. This potential is presented in the left panel in Fig. \ref{V1-sch-wsq} for some values of $\mu$. We have the same static structures $V_{sch}(x)$ for the antikink $\phi_{\bar K}(x)$. From the figure, we note for all range $\mu>0$ there is the possibility of occurrence of bound states. For small values of $\mu$ the potential has a global minimum at $x=0$. With the increasing of $\mu$, the potential becomes higher. For $\mu > 1.3$ the point $x=0$ turns to be a local maximum and begins to appear two minima in the potential, as one can see for $\mu=2$. For even larger parameter values, the asymptotic value of $V_{sch}$ increases, whereas the width of potential decreases.  

As we remarked in the previous section, the vibrational modes are important for the understanding of the intricate collision process. We solved the Schr\"odinger-like equation with $V_{sch}$ for several values of the parameter $\mu$. In the right panel of Fig. \ref{V1-sch-wsq} we present our main results for the emergence of bound states. The increasing of $\mu$ leads to the emergence of new bound states. We have the presence of the translational and one vibrational mode for small values of $\mu$. For $ 0.7 \lesssim \mu \lesssim 1.8$, $ 1.8 \lesssim \mu \lesssim 3.8$ and $ \mu \gtrsim 3.8$  we note the presence of, respectively, two, three and four vibrational modes. We observed also that the energy of the bound states increases continuously for large values of $\mu$. 

%%%%%%%%%%%%%%%%%%%%%%%%%%%%%%%%%%%%%%%%%%%%%%%%%%%%%%%%%%%%%%%%%%%%%%
\begin{figure}
		\includegraphics[{angle=0,width=8cm}]{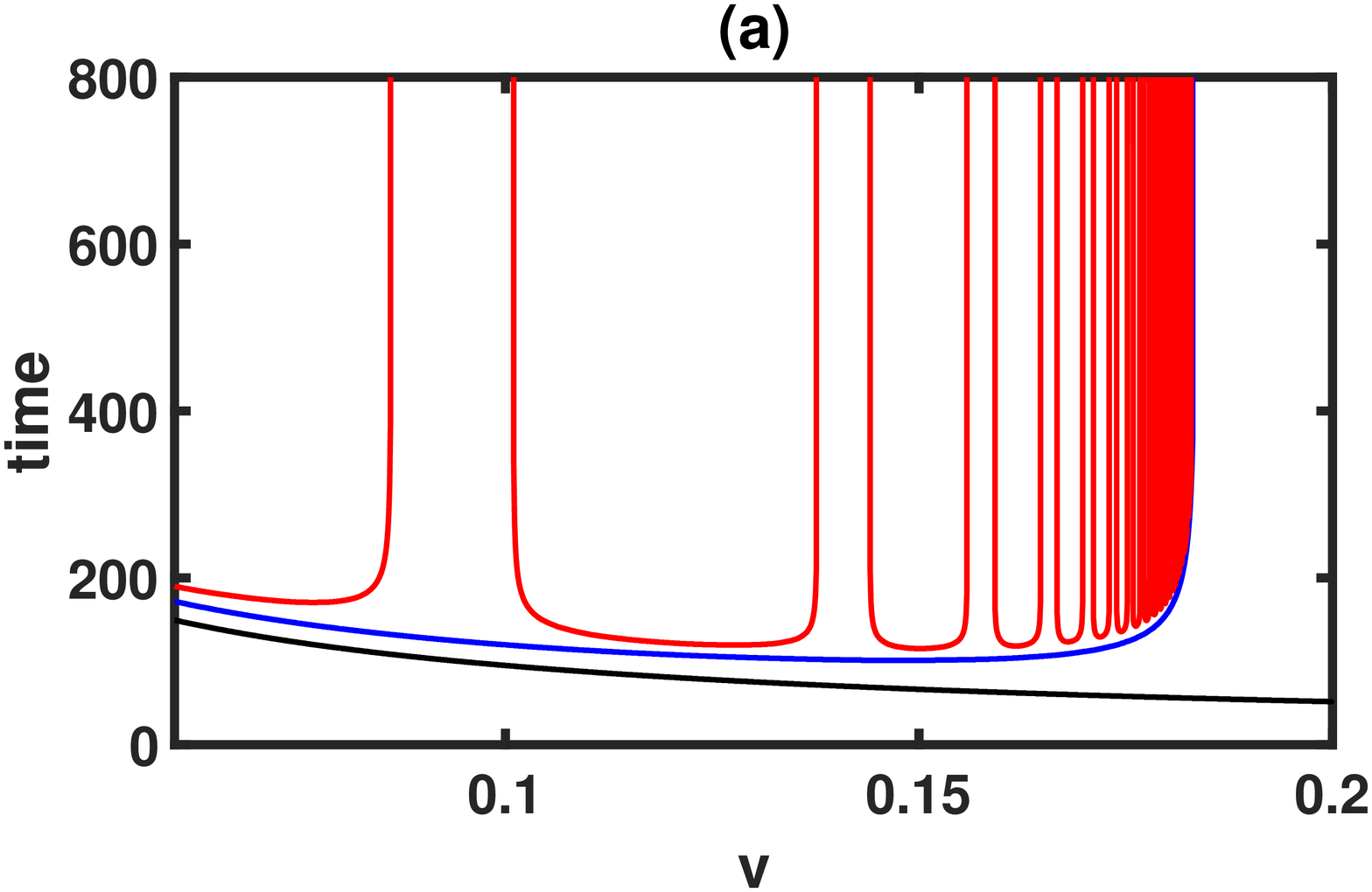} % mu=0.5,height=5cm
		\includegraphics[{angle=0,width=8cm}]{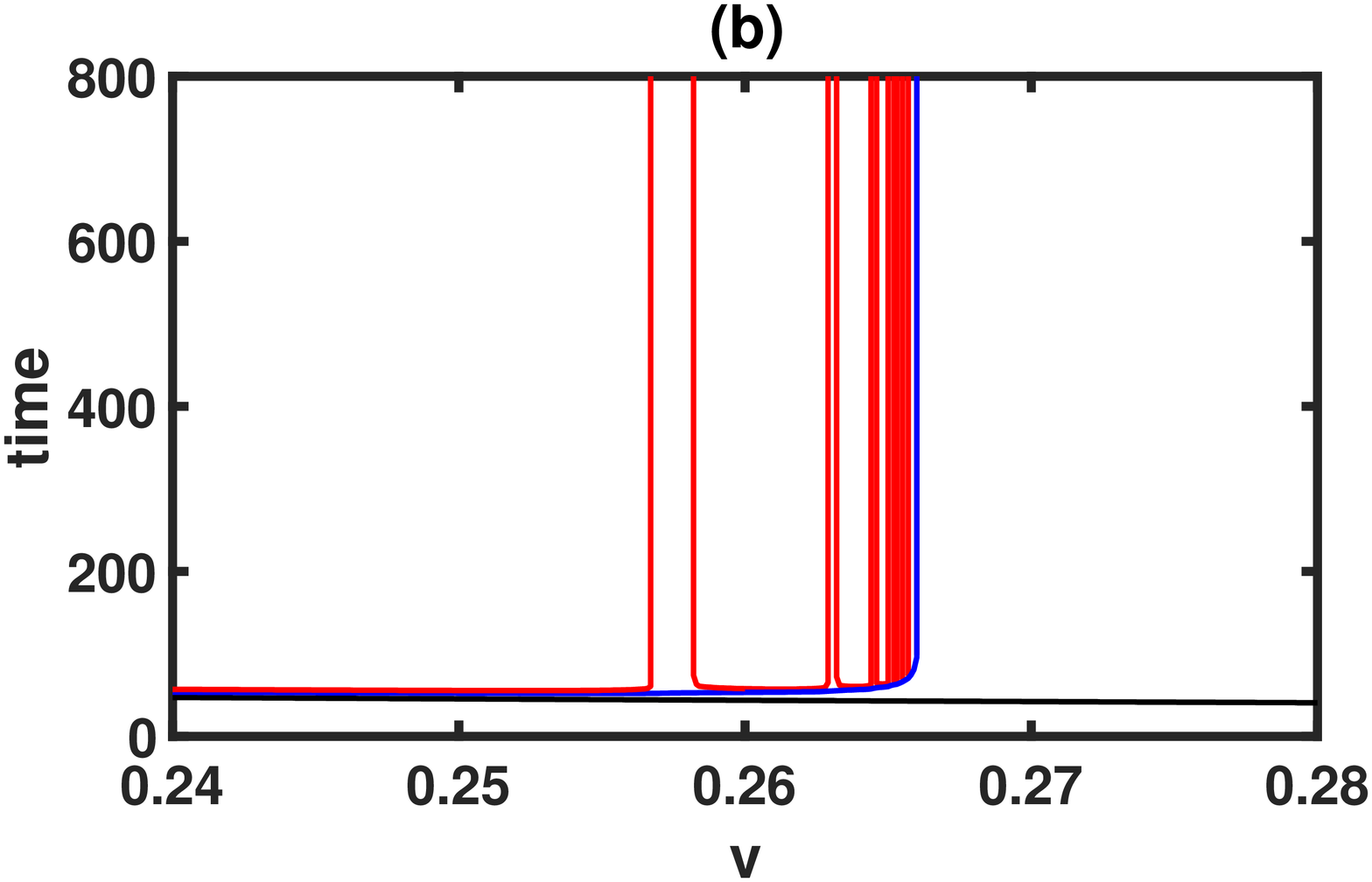} % mu=5.0
		\caption{Kink-antikink collisions: the times to first (black), second (blue) and third (red) bounces for kink-antikink collisions as a function of initial velocity for (a) $\mu=0.5$ and (b) $\mu=5.0$.}
		\label{V1-scatt}
\end{figure}
%%%%%%%%%%%%%%%%%%%%%%%%%%%%%%%%%%%%%%%%%%%%%%%%%%%%%%%%%%%%%%%%%%%%%

The structure of scattering for this model is depicted in the Figs. \ref{V1-scatt}a-b. This figure shows the time $t$ of the first three bounces as a function of initial velocity $v$. The two-bounce windows are visible at the intervals where the time for third kink-antikink collisions diverge. The Fig. \ref{V1-scatt}a correspond the kink-antikink collision for $\mu=0.5$, a case with only one vibrational mode (see also the Fig. \ref{V1-sch-wsq}b). For $v<v_c \sim 0.1836$, bion states are achieved, where the scalar field at the center of mass $\phi(0,t)$ changes after the scattering from the initial value $\phi \simeq 0.96$ to erratic oscillations around the adjacent vacuum $\phi \sim -0.96$. For $v>v_c$ the output is an inelastic scattering between the pair which corresponds to one-bounce around the vacuum $\phi \sim 0.96$. The figure shows the complete set of two-bounce windows without any deformation, and this is related to transfer of energy from translation mode to vibrational mode. Note that the thickness of the two-bounce windows decrease continuously and accumulate around $v_c$.  The Fig. \ref{V1-scatt}b  shows that larger values of $\mu$ contribute to suppress part of the two-bounce windows.  This effect is due to the presence of more than one vibrational state. During the scattering, the energy of the translational mode can be transferred partially to the three vibrational modes. This makes more difficult the realization of the mechanism of resonant energy exchange between the translational and one vibrational mode. Also, from the Fig. \ref{V1-scatt}b we see that  for large values of $\mu$, the critical velocity $v_c$ grows with $\mu$. This  signals a stronger attractive interaction of the kink-antikink pair for larger values of $\mu$. The behavior of behavior of $v_c$ with $\mu$ is not monothonic, showing a minimum around $\mu=1$. 

%%%%%%%%%%%%%%%%%%%%%%%%%%%%%%%%%%%%%%%%%%%%%%%%%%%%%%%%%%%%%%%%%%%%%%
\begin{figure}
	\includegraphics[{angle=0,width=8cm}]{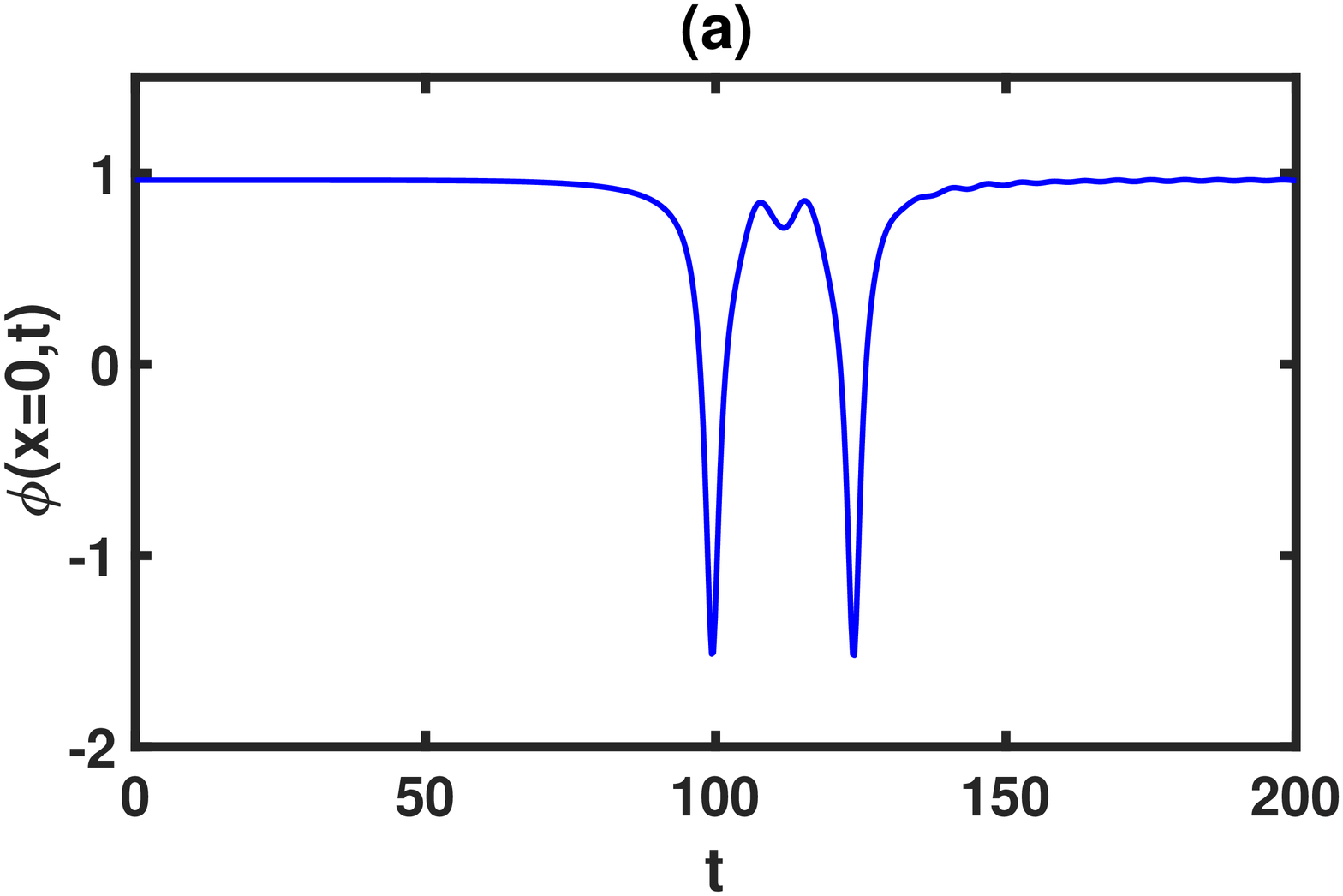}
	\includegraphics[{angle=0,width=8cm}]{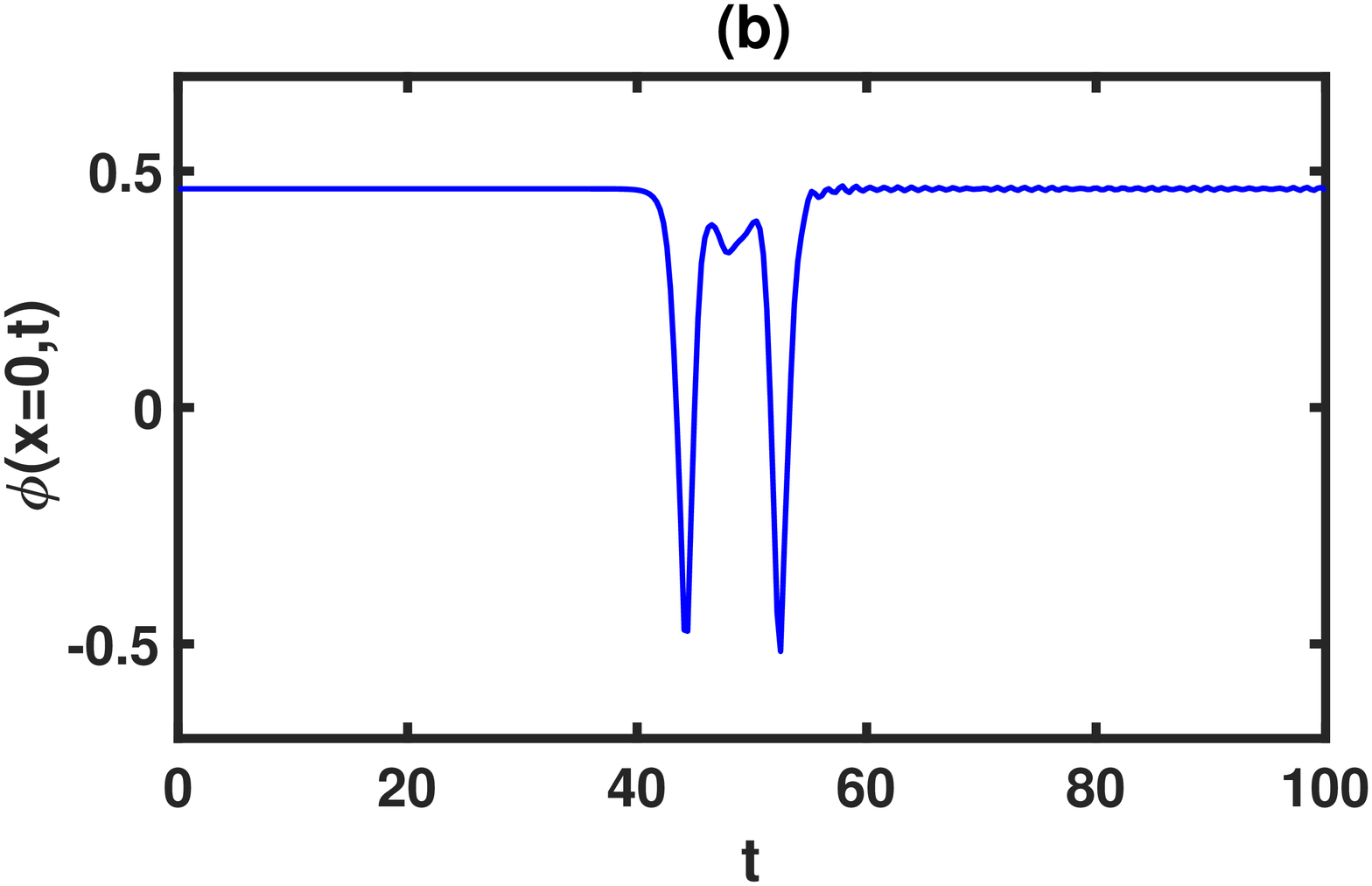}
		\caption{Scalar field at the center of mass $\phi(x=0,t)$ versus time for (a) $\mu=0.5$ with $v=0.0925$ and (b) $\mu=5.0$ with $v=0.2575$.}
	\label{V1-phix0}
\end{figure}
%%%%%%%%%%%%%%%%%%%%%%%%%%%%%%%%%%%%%%%%%%%%%%%%%%%%%%%%%

%%%%%%%%%%%%%%%%%%%%%%%%%%%%%%%%%%%%%%%%%%%%%%%%%%%%%%%%%%%%%%%%%%%%%
\begin{figure}
	\includegraphics[{angle=0,width=8cm}]{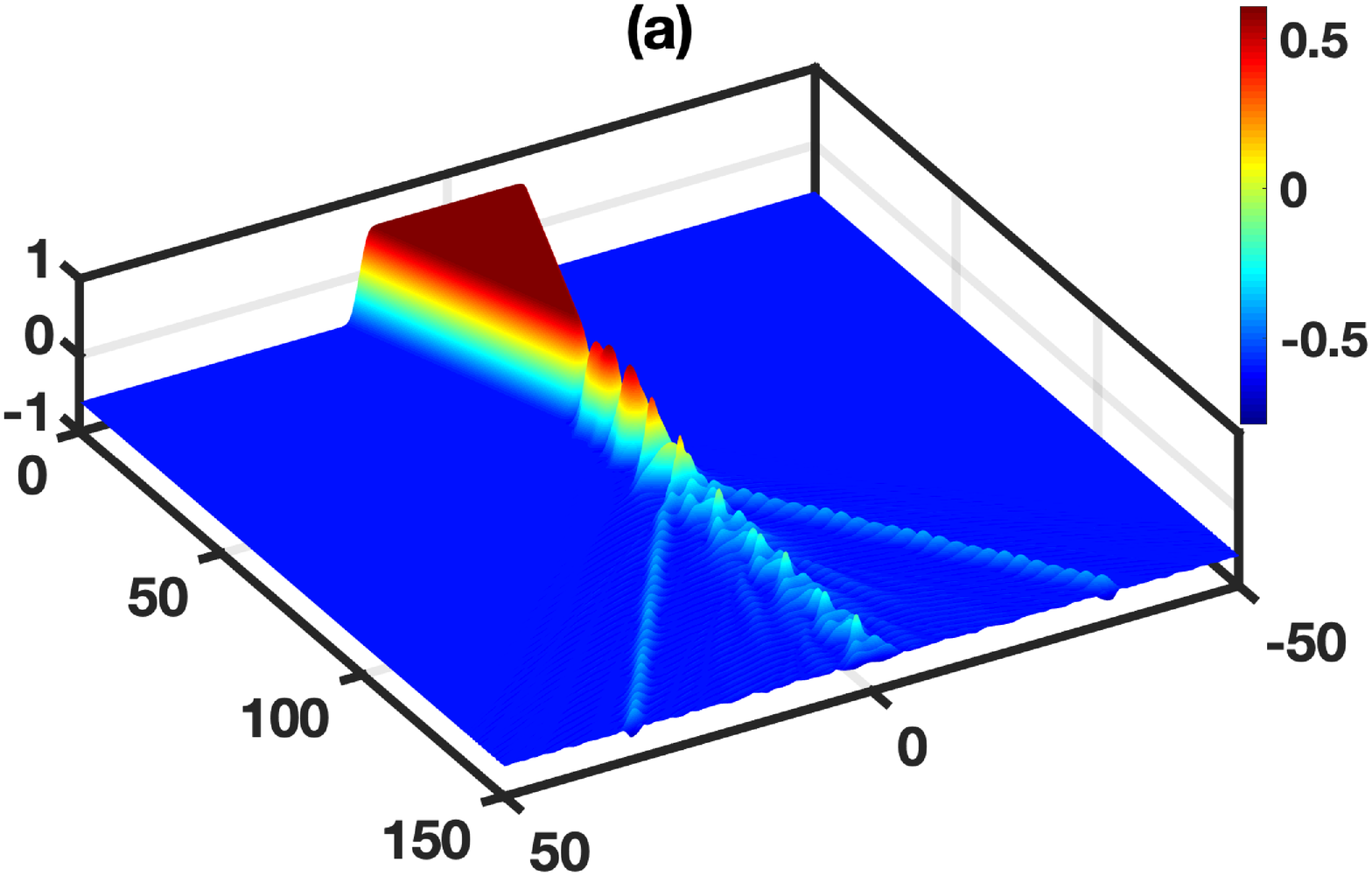} % V1_v_02234_mu_3
		\includegraphics[{angle=0,width=8cm}]{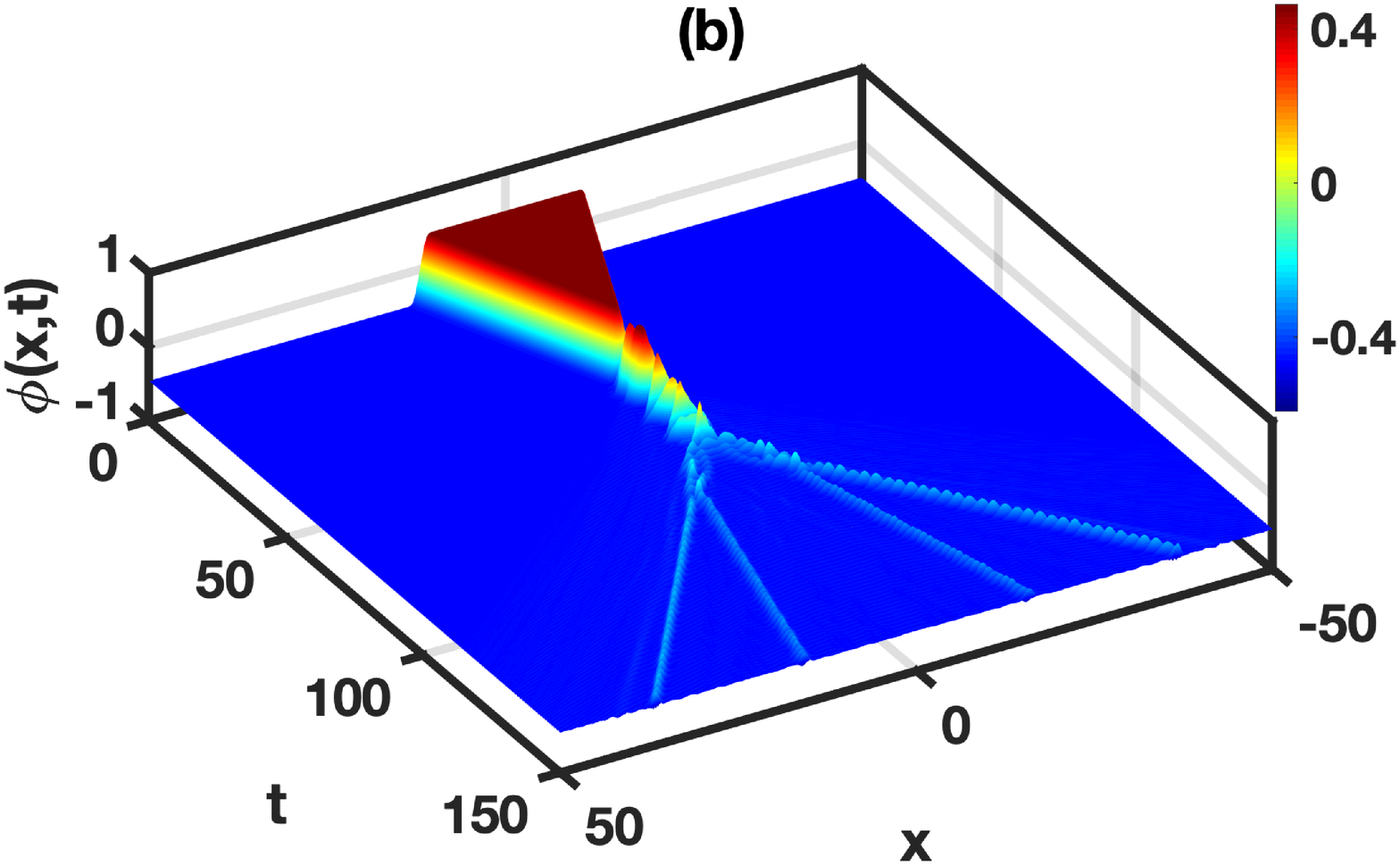} % V1_v_02625_mu_5
	\caption{Kink-antikink collisions and oscillon production  in the model 1:  (a) $\mu=3.0$ and $v=0.2234$,  (b) $\mu=5.0$  and $v=0.2625$.} 
	\label{V1-oscill}
\end{figure}
%%%%%%%%%%%%%%%%%%%%%%%%%%%%%%%%%%%%%%%%%%%%%%%%%%%%%%%%%%%%%%%%%%%%%

One interesting aspect of the kink-antikink scattering in this model is the production of oscillons, that is, the long-lived and low-amplitude oscillation of the scalar field around one vacuum. For low values of $\mu$ there is no formation of such states, and the scattering is restricted to  n-bounces and bion states. For larger values of $\mu$ there is the possibility of oscillon formation, like those depicted in the Figs.  \ref{V1-oscill}a-b. We remark that the number of formed oscillons depends not only on $\mu$, but also on the initial velocity $v$. For example, in the Fig. \ref{V1-oscill}a
we see, for $\mu=3$, the production of two oscillons, whereas for $\mu=5$, we have the production of four oscillons (Fig. \ref{V1-oscill}b). We note that larger values of  the parameter $\mu$ favor the occurrence of more definite oscillon states, having higher harmonicity and correspondingly higher lifetime. 

The absence of oscillons for small values of $\mu$ conforms with the results that for $\mu$ small, the above model becomes the $\phi^4$ model, and that up to now there are no evidence for the presence of oscillons in the scattering of kinks in the $\phi^4$ model.

%%%%%%%%%%%%%%%%%%%%%%%%%%%%%%%%%%%%%%%%%%%%%%%%%%%%%%%%%%%%%%%%%%%%%%%%%%%%%%%%%
\subsection{Model 2}

Let us consider another potential \cite{dk2}
\begin{eqnarray}
V_2(\phi)=\frac{\mu^2}{8 \alpha} \bigg( \frac{\sinh^2(\arcsinh(\mu) \phi)}{\mu^2} - 1 \bigg)^2,
\label{pot2}
\end{eqnarray}
where $\alpha = \Big( (\mu^2+1)\arcsinh^2(\mu) \Big)^{-1}$ and $\mu$ is the deformability parameter.

%%%%%%%%%%%%%%%%%%%%%%%%%%%%%%%%%%%%%%%%%%%%%%%%%%%%%%%%%%%%%%%%%%%%%
\begin{figure}
	\includegraphics[{angle=0,width=8cm}]{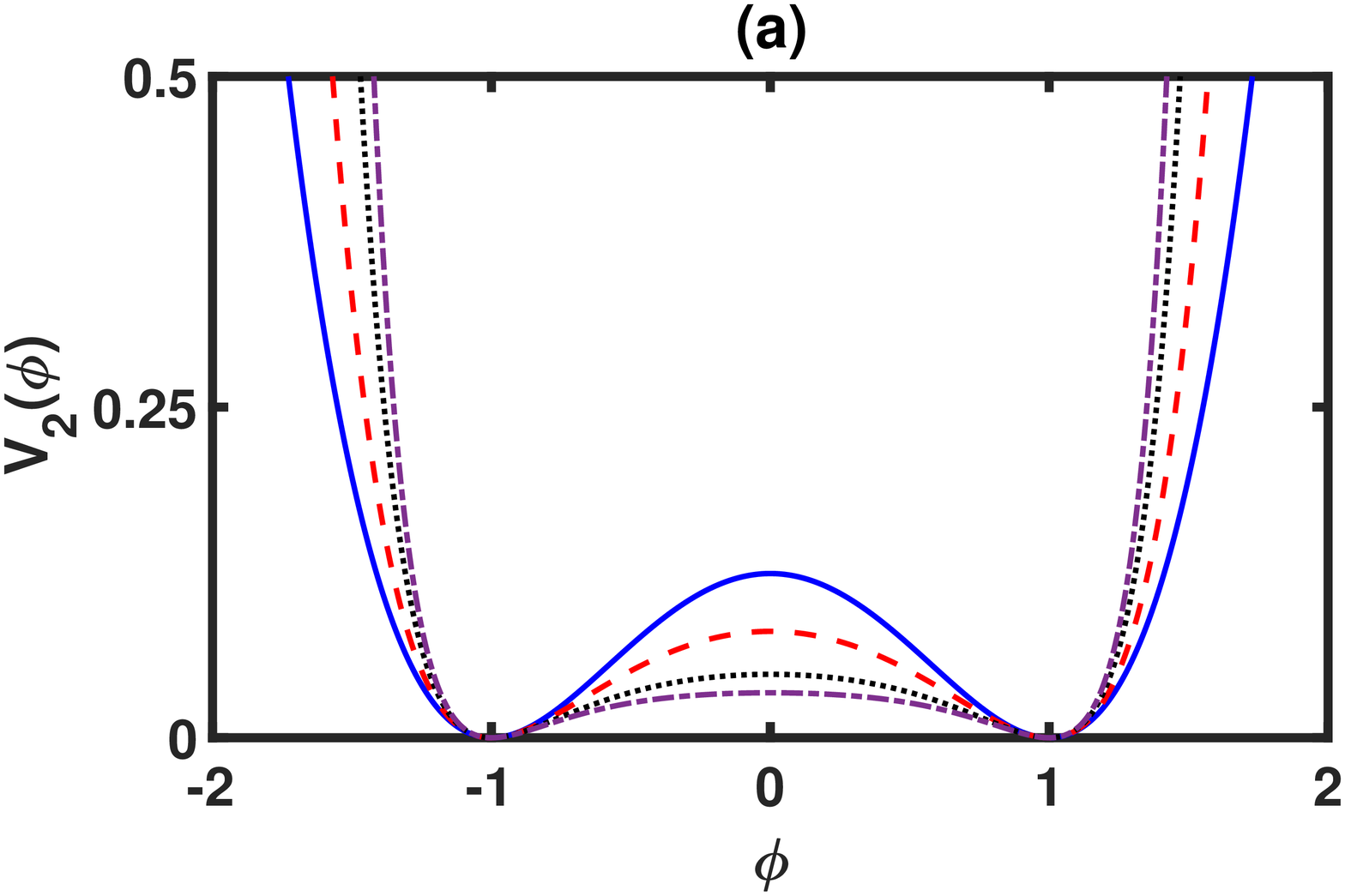}
	\includegraphics[{angle=0,width=8cm}]{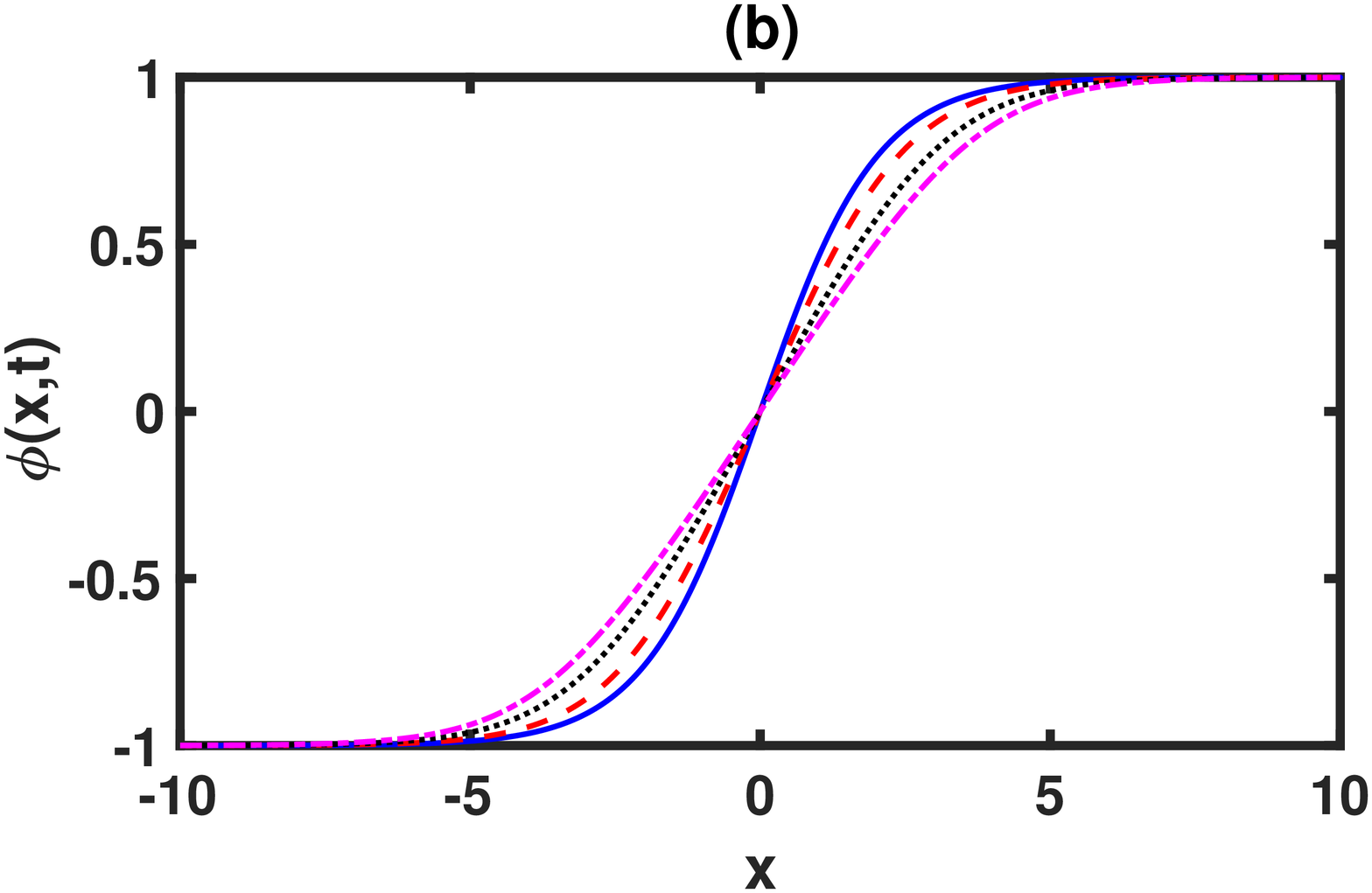}
	\includegraphics[{angle=0,width=8cm}]{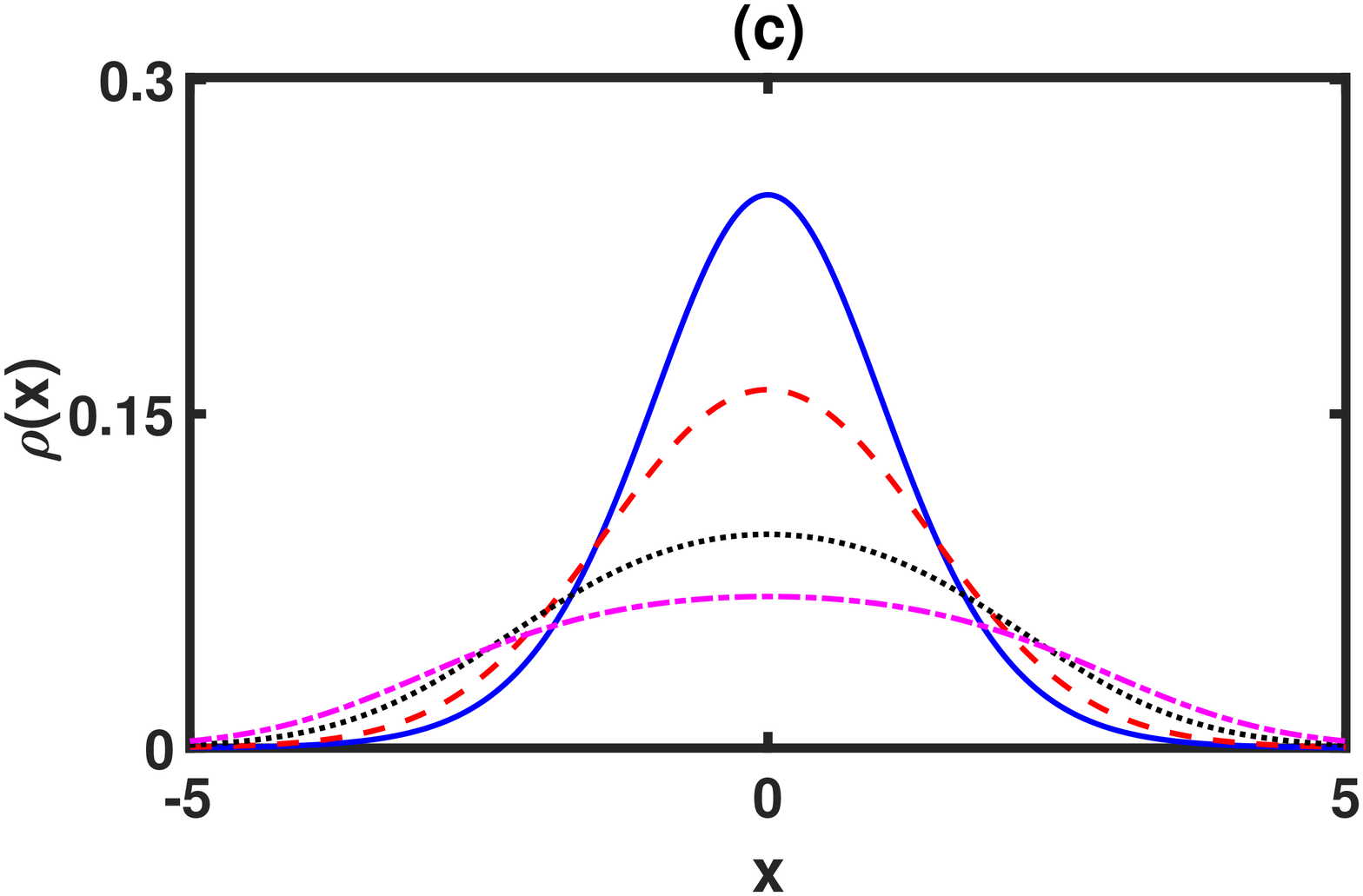}
	\includegraphics[{angle=0,height=5cm,width=8cm}]{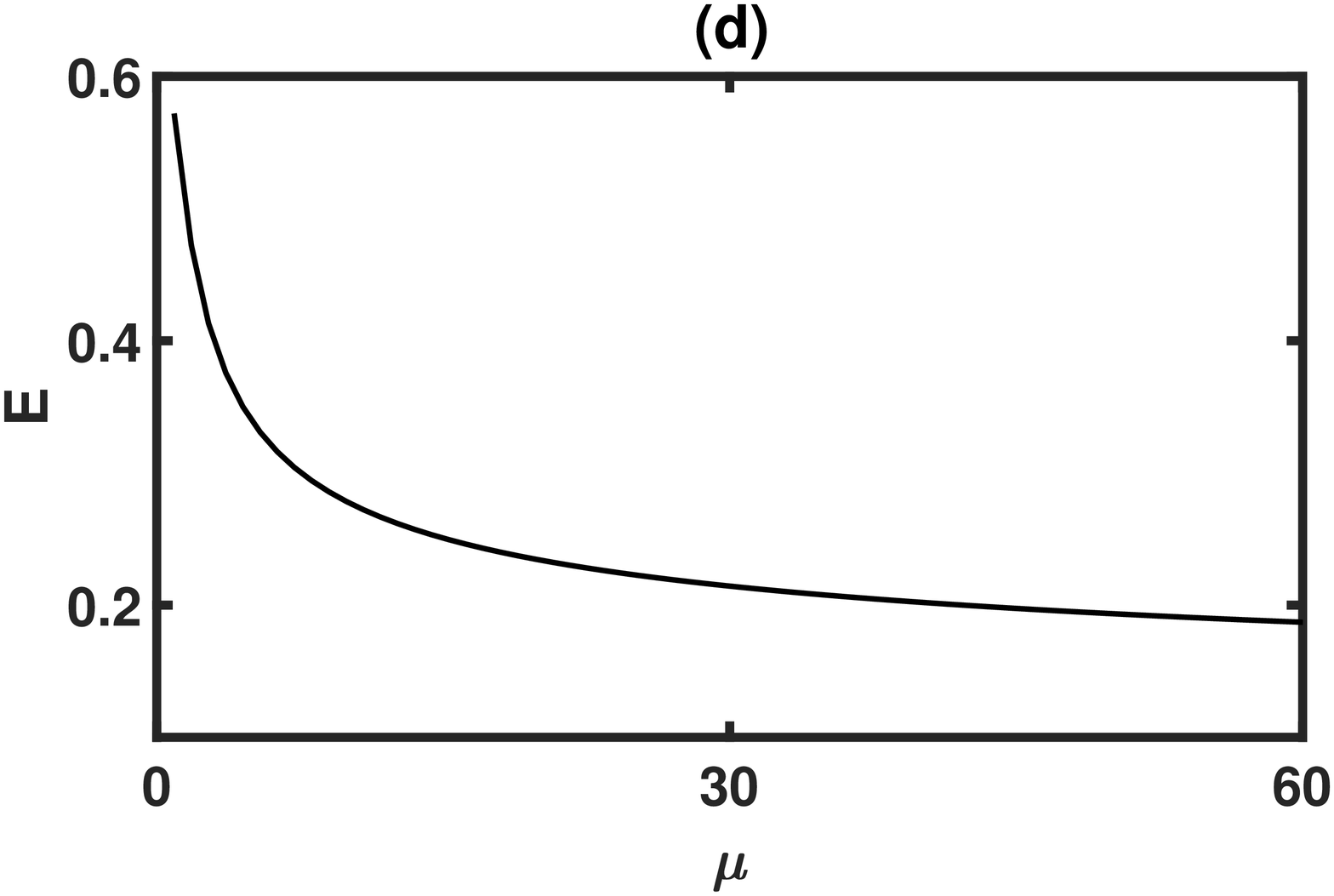}
	\caption{Model 2: (a) Potential $V_2(\phi)$, (b) field $\phi(x)$ and (c) energy density $\rho(x)$  for fixed $\mu=0.1$ (blue solid), $\mu=1$ (red dash), $\mu=2$ (black dotted) and $\mu=3$ (purpple dash-dotted) and (d) the kink energy or rest mass as function of $\mu$.}
	\label{V2-phi-dens}
\end{figure}
%%%%%%%%%%%%%%%%%%%%%%%%%%%%%%%%%%%%%%%%%%%%%%%%%%%%%%%%%%%%%%%%%%%%%

The Fig. \ref{V2-phi-dens}a shows that the potential has two minima in $\phi=\pm 1$ and one local maximum at the origin and that barrier height decreases with $\mu$. Compare with the Fig. \ref{V1-phi-dens}a for the potential $V_1$, where the barrier height is constant. The static kink solution is given by \cite{dk1}
\begin{eqnarray}
\phi_K(x) = \frac{1}{\arcsinh(\mu)}\arctanh\Bigg( \frac{\mu}{\sqrt{1+\mu^2}} \tanh\bigg( \frac{x}{2} \bigg) \Bigg).
\label{sol2}
\end{eqnarray}
The model here also supports first order equations, now given by
\be
\frac{d\phi}{dx}=\pm \frac{dW_2}{d\phi} = \pm \frac\mu{2\sqrt{\alpha}} \biggl(  \frac{\sinh(\arcsinh^2(\mu\phi))}{\mu^2}-1 \biggr).
\ee
The kink and antikink of this model obeys these first order equations, so they are also linearly stable.

%%%%%%%%%%%%%%%%%%%%%%%%%%%%%%%%%%%%%%%%%%%%%%%%%%%%%%%%%%%%%%%%%%%%%
\begin{figure}
	\includegraphics[{angle=0,width=8cm}]{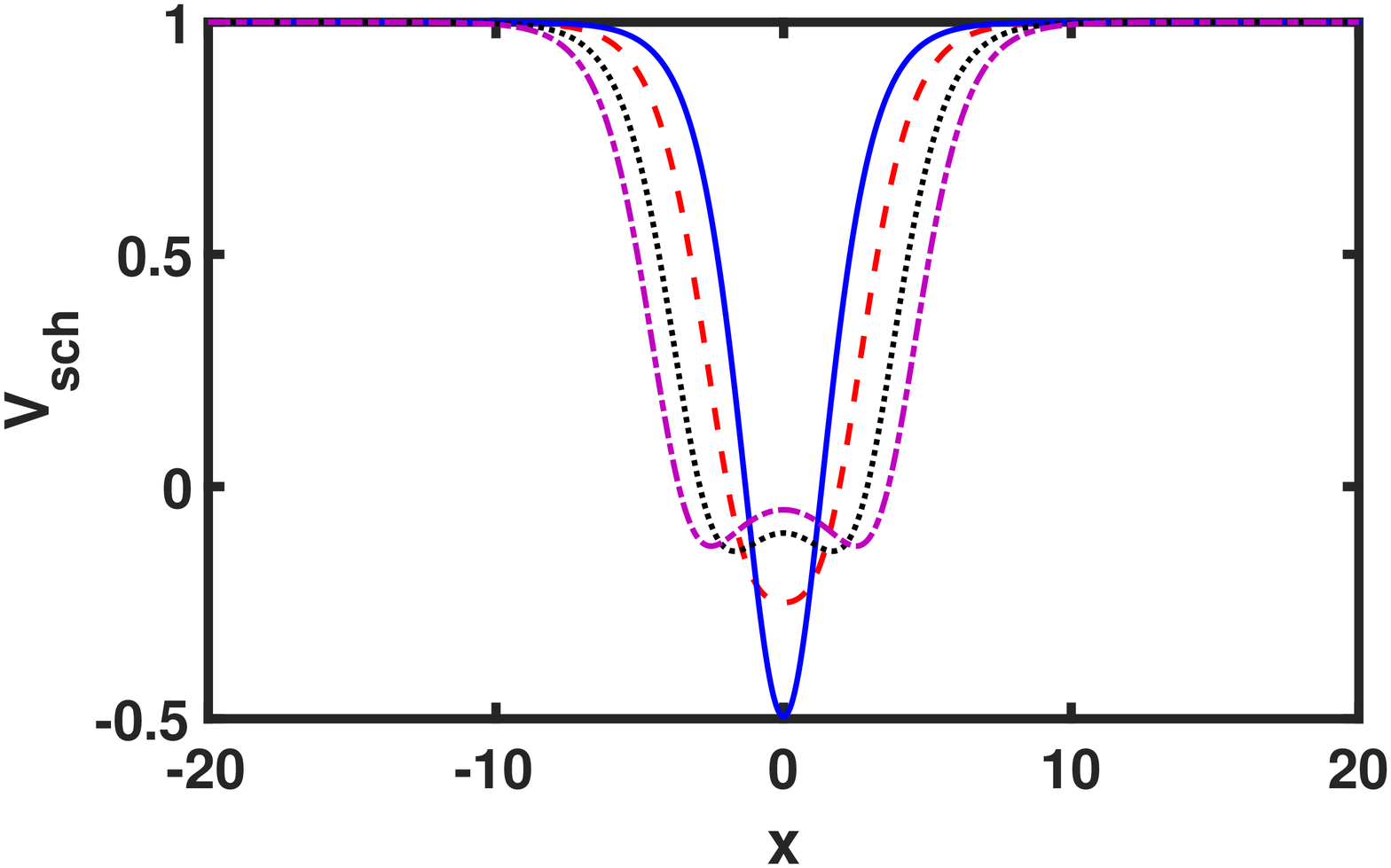}
\includegraphics[{angle=0,width=8cm}]{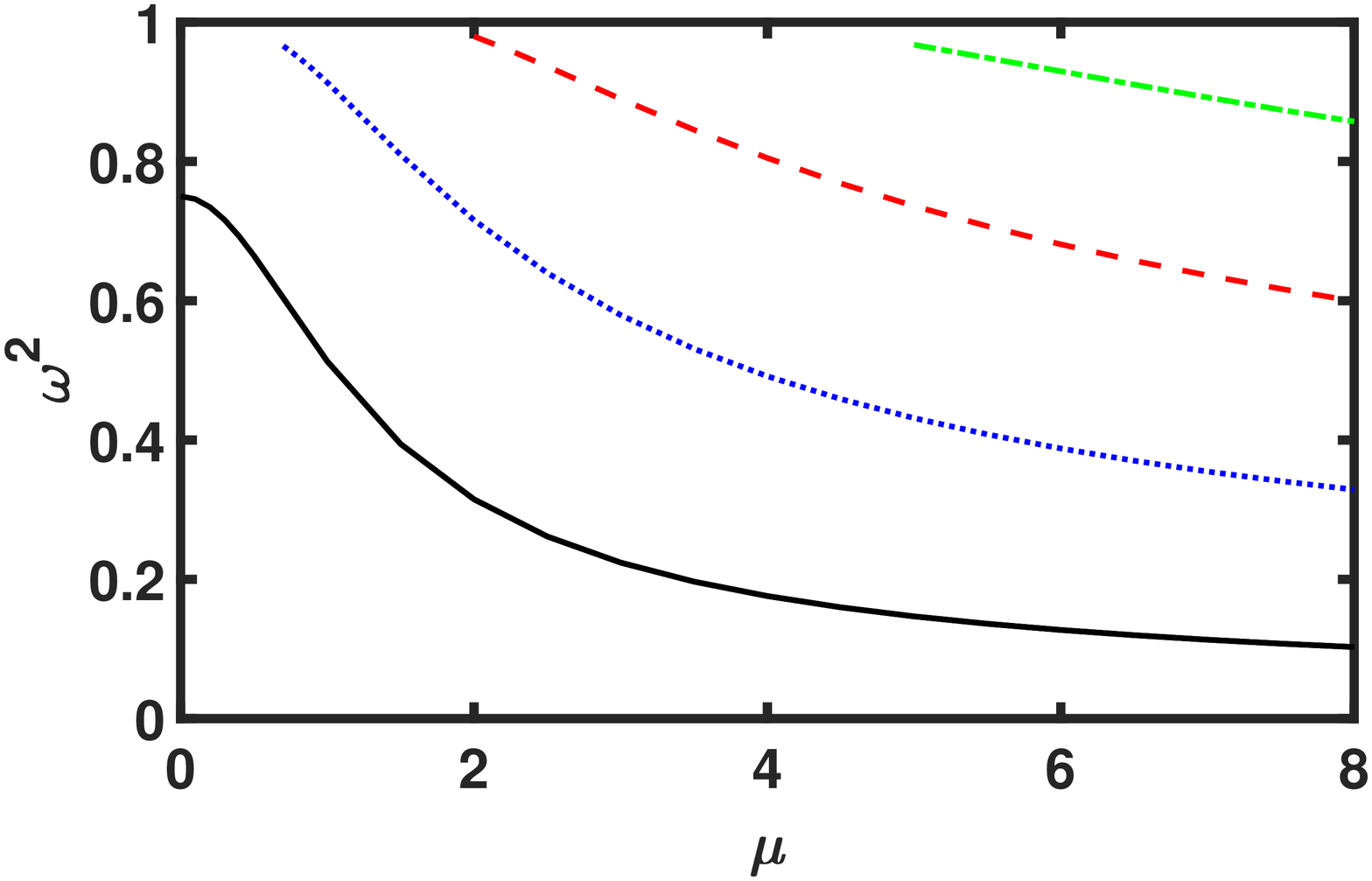}
	\caption{ (Left) The Schr\"odinger-like potential $V_{sch}$ for fixed $\mu=0.1$ (blue solid), $\mu=1$ (red dash), $\mu=2$ (black dotted) and $\mu=3$ (purpple dash-dotted). (Right) The squared frequencies $\omega^2$ of the vibrational states as a function of the parameter $\mu$.}
	\label{V2-sch-wsq}
\end{figure}
%%%%%%%%%%%%%%%%%%%%%%%%%%%%%%%%%%%%%%%%%%%%%%%%%%%%%%%%%%%%%%%%%%%%%

The Figs. \ref{V2-phi-dens}b depicts some plots of the scalar field profile $\phi(x)$ for some values of $\mu$. Note from the figure that the minima are independent of $\mu$ ($\phi=\pm 1$). The Fig. \ref{V2-phi-dens}c shows that the energy density is a localized function around $x=0$. We note that its height decreases with $\mu$ (compare with the Fig.  \ref{V1-phi-dens}c for the energy density for the model $V_1$, where this is constant). Also, we see that its thickness grows with $\mu$ (compare again with the Fig.  \ref{V1-phi-dens}c for the model $V_1$, where this characteristic decreases with $\mu$).  In the Fig. \ref{V2-phi-dens}d. we see that the kink rest energy for this second model also decreases with the increasing of $\mu$.  However, this decreasing occurs in a lower rate in comparison to the $V_1$ model (compare with the Fig.  \ref{V1-phi-dens}d).

The stability potential for the kink in this model is given by 
\be
V_{sch}(x)=\frac{\mu^2\tanh^4\big(\frac x2 \big) + 3\tanh^2\big(\frac x2 \big) - \mu^2 - 1}{2 \bigg( \mu^2\tanh^2\big( \frac{x}{2} \big) - \mu^2 - 1 \bigg)^2}.
\ee
Some plots of $V_{sch}$ are presented in the Fig. \ref{V2-sch-wsq}a for some values of $\mu$. This figure shows that  there is always a possibility of occurrence of bound states. For small values of $0<\mu\lesssim 1.4$ we observe the minimum of potential centered at $x=0$. For larger values of $\mu$, the point $x=0$ turns to be a local maximum with the appearance of two minima in the potential. The increasing of $\mu$ reduces the depth of $V_{sch}$, however the width of potential increases. Moreover, the asymptotic value of $V_{sch}$ is independent of $\mu$ in this case.

%%%%%%%%%%%%%%%%%%%%%%%%%%%%%%%%%%%%%%%%%%%%%%%%%%%%%%%%%%%%%%%%%%%%%
\begin{figure}
	\includegraphics[{angle=0,width=8cm,height=4cm}]{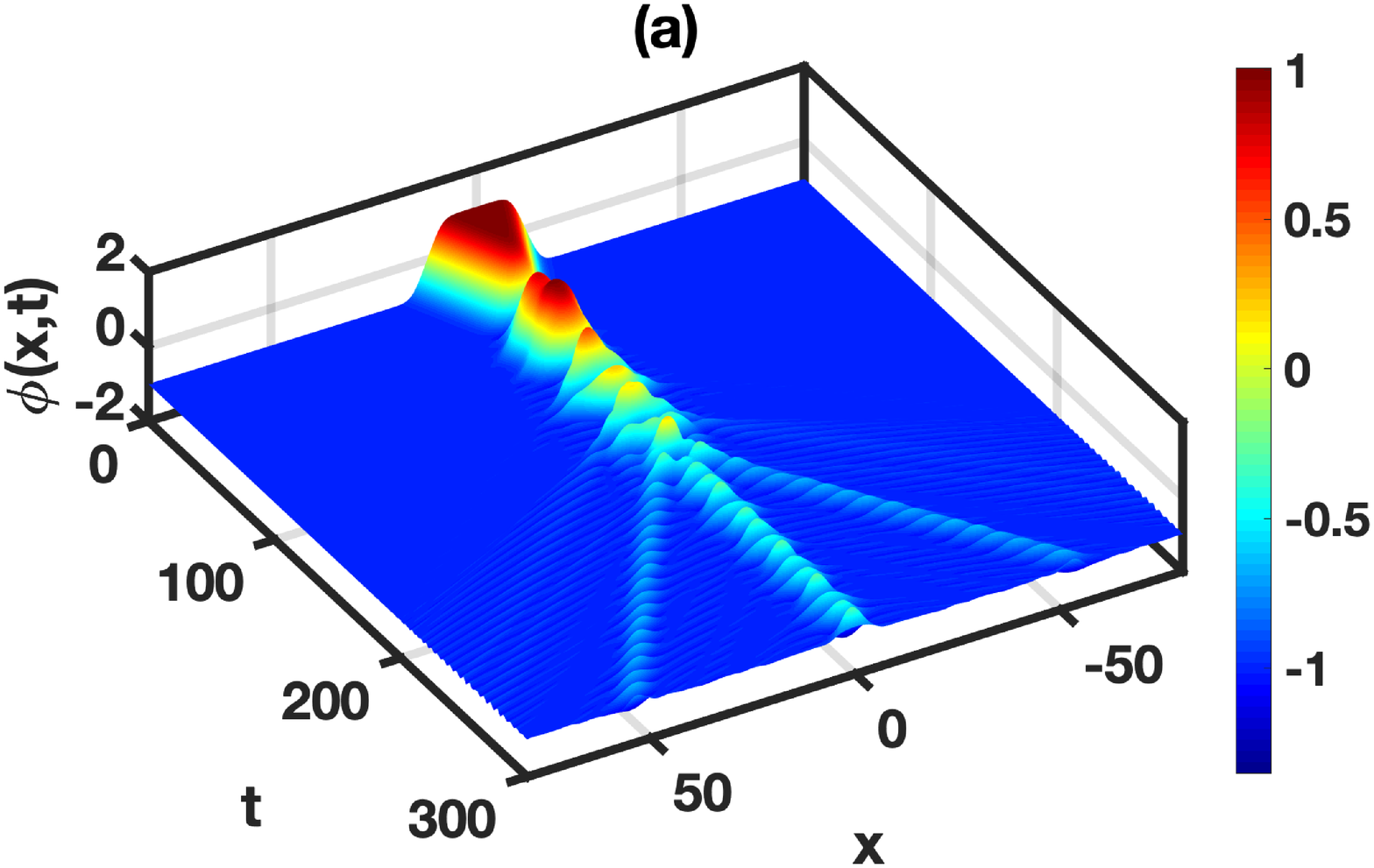} %V2_v_02303_mu_3
	\includegraphics[{angle=0,width=8cm,height=4cm}]{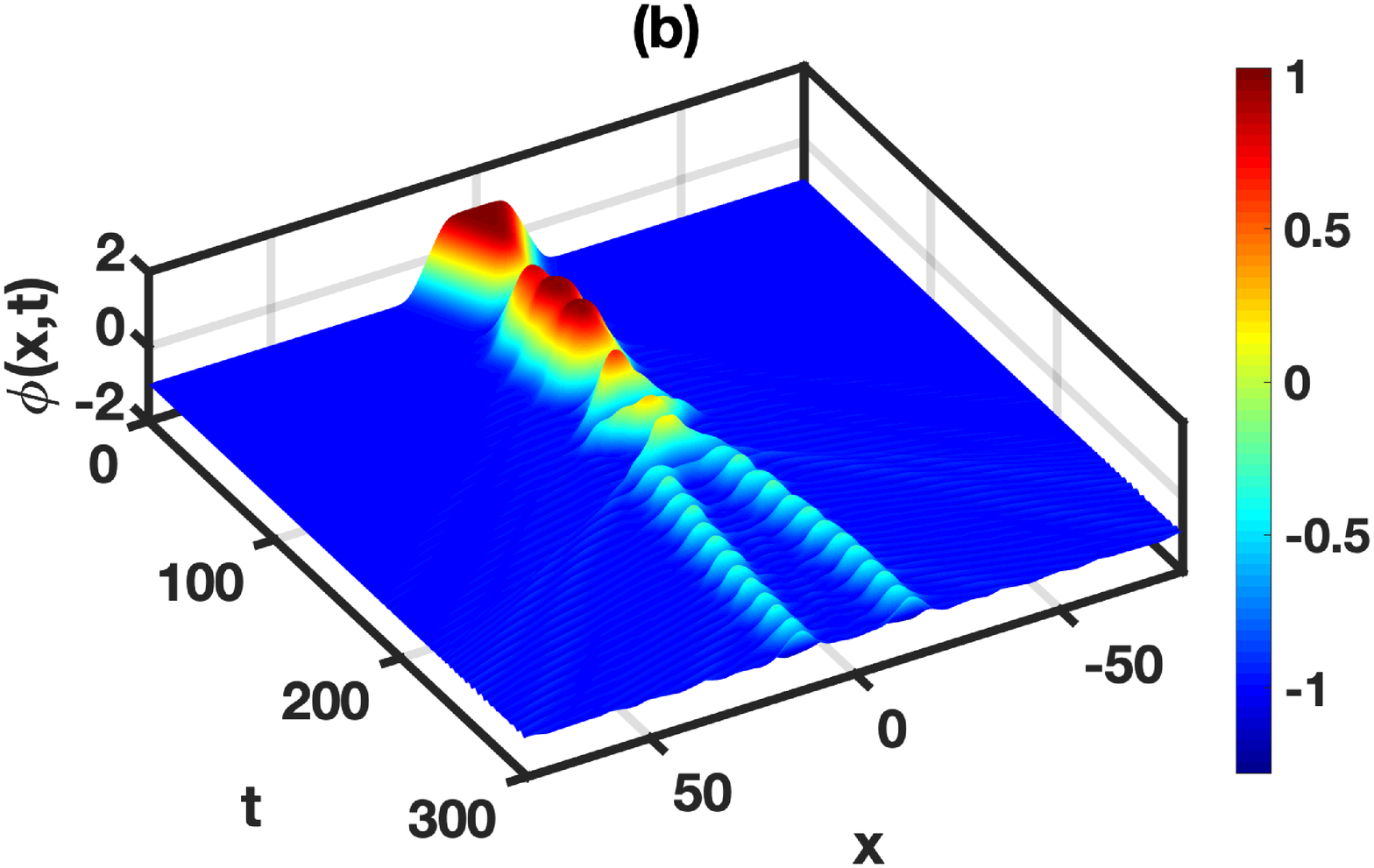} %V2_v_0264_mu_5		
	\caption{Kink-antikink collisions and oscillon production in the model 2: (a) $\mu=3.0$  and $v=0.2303$, (b)  $\mu=5.0$ and $v=0.264$.}
	\label{V2-oscill}
\end{figure}
%%%%%%%%%%%%%%%%%%%%%%%%%%%%%%%%%%%%%%%%%%%%%%%%%%%%%%%%%%%%%%%%%%%%%

We solved the corresponding Schr\"odinger-like equation for several values of $\mu$ and the structure of vibrational modes is summarized in the Fig. \ref{V2-sch-wsq}b. For all values of $\mu$ there is always a zero-mode. The increasing the $\mu$ leads to the emergence of new bound states. In particular, for $\mu \geq 0.7 $ we note the presence of extra bound state. When $\mu \geq 2.0$ and $\mu \geq 5.0$ we observe the third and fourth vibrational states. There one can see that energy of extra bound states decreases with $\mu$. Compare with the Fig. \ref{V1-sch-wsq}b for the model 1, where $\omega^2$ grows with $\mu$.

The pattern of kink-antikink scattering for the $V_2$ model is so close to the observed by $V_1$ that we do not need to show the results here. That is, the suppression of two-bounce windows with $\mu$ is very similar to the Fig. \ref{V1-scatt}. One also has the plot of  $v_c$ with $\mu$ with a minimum around $\mu\sim 1$. Remember that in the model $V_2$, the height of the barrier is changed, but the position of degenerate minima is the same. In the model $V_1$, on the contrary, the height of the barrier is fixed, whereas the position of degenerate minima changes. Our numerical investigation  showed that both scenarios are almost indistinguishable with respect to the scattering process. 

The kink-antikink scattering in the $V_2(\phi)$ model also showed the production of oscillons. In the Fig. \ref{V2-oscill}a and \ref{V2-oscill}b we see clearly three and two escaping oscillons for $\mu=3$. The same results happen for the Fig. \ref{V2-oscill}c and \ref{V2-oscill}d with $\mu=5$. The appearance of oscillons is more favorable with the presence of the extra vibrational state. In particular, for low values of $\mu$ there is no formation of such states. This is in accordance with the facts that for small values of $\mu$ the model approaches the $\phi^4$ model, and that up to now there are no evidence for the presence of oscillons in collisions of kinks in the $\phi^4$ model.

%%%%%%%%%%%%%%%%%%%%
\section { Conclusions  }

%%%%%%%%%%%%%%%%%%%%%%%%%%%%%%%%%%%%%%%%%%%%%%%%%%%%%%%%%%%%%%%%%%%%%%%%%%%%%%%

We have analyzed two models of potentials with two degenerated minima, with interest in the two effects: the difference between the minima and the height of the barrier of the potential. For the models $V_1$ and $V_2$ one of the characteristics is fixed, whereas the other varies monotonically with the parameter $\mu$. We found that the dynamics of scattering at the center of mass is roughly the same, with the expected one-bounce, bion and two-bounce states. The increasing of $\mu$ is accompanied by the generation of more vibrational states. The suppression of two-bounce windows is due to a kind of destructive interference between the two vibrational modes that forbids the realisation of the resonance mechanism of transferring of energy from the translational mode to the vibrational mode. This effect of suppression was already described in other models \cite{sgno} and is here also identified.

We have also observed the production of oscillons, that is, long-lived states that oscillate around one trivial minima of the potential. One sees that the production of oscillons is favored in the following situations: i) for the model $V_1$, for fixed barrier, when the difference between the two minima is smaller; ii) for the model $V_2$, for fixed minima, when the barrier is smaller. For both  $V_1$ and $V_2$ models, the unifying factor that favors the production of oscillons is the energy or rest mass $E$ of the kink. Indeed, in both models, this quantity decreases monotonically with $\mu$. 

In addition, another interesting phenomenon was shown in the kink-antikink scattering of the two models $V_1(\phi)$ and $V_2(\phi)$. As we can see in the Figs. \ref{V1-oscill} and \ref{V2-oscill}, after some collisions, long-lived, quasi-harmonic and low amplitude oscillating structures are formed and escape to infinity. These states, known as oscillons, can occur for $v<v_c$. The appearance of oscillons are extremely sensitive to the initial velocity. Despite this, we have identified some determinant factors for the production of oscillons.  Fist of all, one sees that the production of oscillons is favored in the following situations: i) for the model $V_1$, for fixed barrier, when the difference between the two minima is smaller; ii) for the model $V_2$, for fixed minima, when the barrier is smaller. For both  $V_1$ and $V_2$ models, the unifying factor that favors the production of oscillons is the lower energy or rest mass $E$ of the kink. Indeed, in both models, this quantity decreases monotonically with $\mu$.  The second aspect to be noted  is that we have not observed the production of oscillons for low values of $\mu$, where the models have only one vibrational state. When the parameter
$\mu$ grows, the increasing in the number of vibrational states results in a greater complexity of the energy distribution of the initial translational modes of the kink-antikink pair, increasing the possibility of production of oscillons.  Also, comparing the results from both potentials, we see that the oscillons with larger amplitudes, but more deformed, are favored for the model $V_2$. Since the harmonicity and propagation without distortion are desirable properties, this signals that model $V_1$, characterized by a fixed height barrier, is more effective for the production of these long-lived states.

\section{Acknowledgements}
A.R.G, F.C.S. and K.Z.N. thank FAPEMA, Funda\c c\~ao de Amparo \`a Pesquisa e ao Desenvolvimento do Maranh\~ao, through grants PRONEX $01452/14$, PRONEM $01852/14$, Universal-$01061/17$, Universal-$01332/17$, Universal-$01191/16$, Universal-$01441/18$. A.R.G thanks CNPq (brazilian agency) through grants $309842/2015$-8, $311501/2018$-4, $437923/2018$-5 for financial support. This study was also financed in part by the Coordena\c c\~ao de Aperfei\c coamento  de Pessoal de N\' ivel Superior - Brasil (CAPES) - Finance Code 001.
D.B. thanks CNPq (grant 306614/2014-6) and Paraiba State Research Foundation (FAPESQ-PB, Grant 0015/2019) for financial support.
%%%%%%%%%%%%%%%%%%%%%%%%%%%%%%%%%%%%%%%%%%%%%%%%%%%%%%%%%%%%%%%%%%%%%%%%%%%%%%%%%

\end{document}